\newif\ifstandardtemplate \standardtemplatetrue
\newif\ifelseviertemplate \elseviertemplatefalse
\newif\ifspringertemplate \springertemplatefalse
\newif\ifwileytemplate    \wileytemplatefalse
\newcommand{\mypackages}{%
  \usepackage{amssymb}
  \usepackage{amsmath}
  \usepackage{amsthm}
  \usepackage[colorlinks=true,allcolors=red]{hyperref}
  \usepackage{graphicx}
  \usepackage{enumitem}
  \usepackage{subfigure}
\usepackage{xfrac}
\usepackage{siunitx}
\graphicspath{{Figures/}{figures/}}
}

\newcommand{\mytitle}{Analysis and design of bistable and thermally reversible metamaterials inspired by shape-memory alloys}
\newcommand{\myabstract}{In this work, we study lattice structures that exhibit a \emph{bistable}
behavior, i.~e., they can snap from one stable state to another, and are also completely
\emph{reversible}, capable of reverting back to its original state through a heat
treatment. We \emph{design} this behavior by constructing lattice structures using networks of
nonlinear springs that display tension-compression asymmetry and have different thermal expansion
coefficients. The mismatch in the thermal expansion coefficients induces residual stresses in the
springs which results in the lattice structure exhibiting bistability at low temperatures and
monostability at high temperatures. This behavior mimics the crystallographic phase transformations
of shape memory alloys, but here artificially introduced in a structural lattice.
By analyzing a representative unit cell, we quantify the effect that the stiffness
and the thermal expansion coefficient of the springs have on the stability of the structural lattice. In
addition, for simple 2D lattices, using the concept of universal unfoldings of singularity theory, we perform a perturbation
analysis to identify the key variables of the structure where controlling defects is important, as
they lead to drastic changes in the bifurcation behavior of the lattice. Finally, we verify
numerically our analytical predictions in both 2D and 3D simulations using continuation
techniques. The examples proposed confirm that the bistable and reversible features of the unit cell
carry on to the macroscale, opening the route for the design of lattice structures for energy
absorption applications that can \emph{heal} with a heat treatment.}

\newcommand{\myack}{The authors gratefully acknowledge the funding received from project
DPI2017-92526-EXP financed by the Spanish Ministry of Economy and Competitiveness.}

\ifspringertemplate
\documentclass[smallcondensed]{svjour3}
\usepackage[final]{changes}
\usepackage{amssymb}
\usepackage{amsmath}
\usepackage{graphicx}
\usepackage{enumitem}

\title{\mytitle%
\thanks{the thanks}}
\author{\ldots \and Ignacio Romero \and \ldots}
\journalname{The journal name}

\institute{I. Romero \at
  IMDEA Materials Institute, Eric Kandel 2, Tecnogetafe, Madrid 28906, Spain\\
  Universidad Polit\'ecnica de Madrid, Jos\'e Guti\'errez Abascal, 2, Madrid 29006, Spain\\
  \email{ignacio.romero@imdea.org}
  \and
  XX \at
  XXX\\
  \email{XXX}}

\titlerunning{\mytitile}
\authorrunning{I. Romero}

\date{Received: date / Accepted: date}

\begin{document}
\maketitle
\begin{abstract}
  \myabstract
\end{abstract}
\fi

\ifelseviertemplate
\documentclass[3p,11pt]{elsarticle}
\mypackages
\newcommand{\mybibstyle}{apalike}

%
\begin{document}
\begin{frontmatter}

\title{\mytitle}

\author{Aditya Vasudevan$^{1}$}
\ead{aditya.vasudevan@imdea.org}

\author{Jos\'e A. Rodr\'{\i}guez-Mart\'{\i}nez$^2$}
\ead{jarmarti@ing.uc3m.es}

\author{Ignacio Romero\corref{cor1}$^{3,1}$}
\ead{ignacio.romero@upm.es}

\address{$^1$IMDEA Materials Institute, Eric Kandel 2, Tecnogetafe, Madrid 28906, Spain}
\address{$^2$Dept.~of Continuum Mechanics and Structural Analysis, University Carlos III of Madrid, Avda. de la Universidad, 30. 28911 Legan\'es, Madrid, Spain}
\address{$^3$Dept.~of Mechanical Engineering, Universidad Polit\'ecnica de Madrid, Jos\'e Guti\'errez Abascal, 2, Madrid 29006, Spain}

\cortext[cor1]{Corresponding author. Universidad Polit\'ecnica de Madrid, Jos\'e Guti\'errez Abascal, 2, Madrid 28006, Spain}

\begin{abstract}
\myabstract
\end{abstract}

\end{frontmatter}

\newenvironment{acknowledgements}{\section*{Acknowledgements}}{}

\fi

\ifwileytemplate
\documentclass[doublespace,times]{nmeauth}
\usepackage[final]{changes}
\usepackage{amssymb}
\usepackage{amsmath}
\usepackage{amsthm}
\usepackage{graphicx}
\usepackage{enumitem}
\usepackage{natbib}
\usepackage{xfrac}
\usepackage{todonotes}
\usepackage{siunitx}
\graphicspath{{Figures/}{figures/}}
\message{Compiling with Wiley template}

\begin{document}
\runningheads{I. Romero}{...}

\title{\mytitle}
\author{Ignacio Romero\affil{1}\affil{2}\corrauth}

\address{\affilnum{1} ETSII, Universidad Politï¿½cnica de Madrid,
         Jos\'{e} Guti\'{e}rrez Abascal, 2, 28006 Madrid, Spain\break
         \affilnum{2} IMDEA Materials Institute, Eric Kandel 2, 28096 Getafe, Madrid, Spain}

\corraddr{Dpto. de Ingenier\'{\i}a Mec\'{a}nica; E.T.S. Ingenieros Industriales;
Josï¿½ Gutiï¿½rrez Abascal, 2; 28006 Madrid; Spain. Fax (+34) 91 336 3004}

\begin{abstract}
\end{abstract}

\maketitle
\fi

\ifstandardtemplate
\documentclass[10pt,a4paper]{article}
\mypackages
\newcommand{\mybibstyle}{unsrt}
\let\citep\cite

\newtheorem{theorem}    {Theorem}[section]
\newtheorem{lemma}      [theorem]{Lemma}
\newtheorem{corollary}  [theorem]{Corollar}
\newtheorem{proposition}[theorem]{Proposition}
\newtheorem{algorithm}  [theorem]{Algorithm}

\theoremstyle{definition}
\newtheorem{definition} [theorem]{Definition}
\newtheorem{problem}    [theorem]{Problem}

\newtheorem{examplex}[theorem]{$\triangleright\;$Ejemplo}
\newenvironment{example}{\medskip\begin{examplex}}{\hfill$\triangleleft$\end{examplex}}

\theoremstyle{remark}
\newtheorem{remark}{Remark}
\newtheorem{remarks}{Remarks}
\newtheorem{note}{Note}

\message{Compiling with default template}

\title{\mytitle}
\author{A. Vasudevan$^{1}$, J. A. Rodr\'{\i}guez-Mart\'{\i}nez$^2$, I. Romero$^{3,1}$\footnote{Corresponding author \texttt{ignacio.romero@upm.es}}}
\date{
  $^1$IMDEA Materials Institute, Eric Kandel 2, Tecnogetafe, Madrid 28906, Spain
  \\[2ex]
  $^2$Dept.~of Continuum Mechanics and Structural Analysis, University Carlos III of Madrid, Avda. de la Universidad, 30. 28911 Legan\'es, Madrid, Spain
  \\[2ex]
  $^3$Universidad Polit\'ecnica de Madrid,
  Jos\'{e} Guti\'{e}rrez Abascal, 2, 28006 Madrid, Spain\\[2ex]}

\newenvironment{acknowledgements}{\section*{Acknowledgements}}{}

\begin{document}
\maketitle
\fi

\newcommand{\concept}[1]{\textbf{\emph{#1}}}
\newcommand{\defined}{:=}
\newcommand{\dev}{{\mathop{\mathrm{dev}}}}
\renewcommand{\div}{{\mathop{\mathrm{div}}}}
\newcommand{\mbs}[1]{\boldsymbol{#1}}
\newcommand{\mcl}[1]{\mathcal{#1}}
\newcommand{\pairing}[2]{\langle{#1},{#2}\rangle}
\newcommand{\dd}[2]{\frac{\mathrm{d} #1}{\mathrm{d} #2}}
\newcommand{\pd}[2]{\frac{\partial #1}{\partial #2}}
\newcommand{\fd}[2]{\frac{\delta #1}{\delta #2}}
\newcommand{\set}[1]{\left\{#1\right\}}
\newcommand{\trace}{{\mathop{\mathrm{tr}}}}
\newcommand{\uptohere}{\centerline{\textcolor{blue}{\rule{6cm}{0.2cm}}}}
\renewcommand{\vec}[1]{\overline{#1}}
\newcommand{\kin}{\ensuremath{k_{\rm{in}}}}
\newcommand{\kout}{\ensuremath{k_{\rm{out}}}}
\newcommand{\alphain}{\ensuremath{\alpha_{\rm{in}}}}
\newcommand{\alphaout}{\ensuremath{\alpha_{\rm{out}}}}
\let\oldLambda=\Lambda
\renewcommand{\Lambda}{\mathit{\oldLambda}}


%
\section{Introduction}
\label{sec:introduction}

Instabilities play an important role in determining the functional capacity of deformable bodies
since they can lead to catastrophic failure in the case, for example, of buckling or brittle
fracture (see, e.g., \cite{Bazant:10} and \cite{Bigoni:14} for in-depth introductions to the topic). Generally considered as
something to be designed against, recent years have seen a tremendous progress in understanding
these instabilities, allowing a change of paradigm where now instead of avoiding them, they can be
exploited. Aided by advances in manufacturing, designs have been proposed that harness these
instabilities to give structures new and surprising capabilities (see~\cite{Kochmann2017, Holmes2019}
and references therein). In particular, instabilities have been exploited to design
mechanical metamaterials, periodic arrays of representative unit cells that when stacked together
possess drastically different properties than the base material. Examples of such designs include
structures with negative Poisson's ratio induced by buckling \citep{Bertoldi2010, Nicolaou2012,Overvelde2012}, snapping based metamaterials for energy absorption \citep{Shan2015, Rafsanjani2015,
Restrepo2015, Haghpanah2016, Coulais2016, yang2020li, chen2020qn, jamshidian2020mu}, origami or kirigami tessellations with multistability
\citep{Silverberg2014, Filipov2015, Sussman2015, Zhang2015, Overvelde2016}, etc.

\begin{figure}[ht!]  \centering \subfigure[\label{fig:fig_1a}]
{\includegraphics[width=0.45\textwidth]{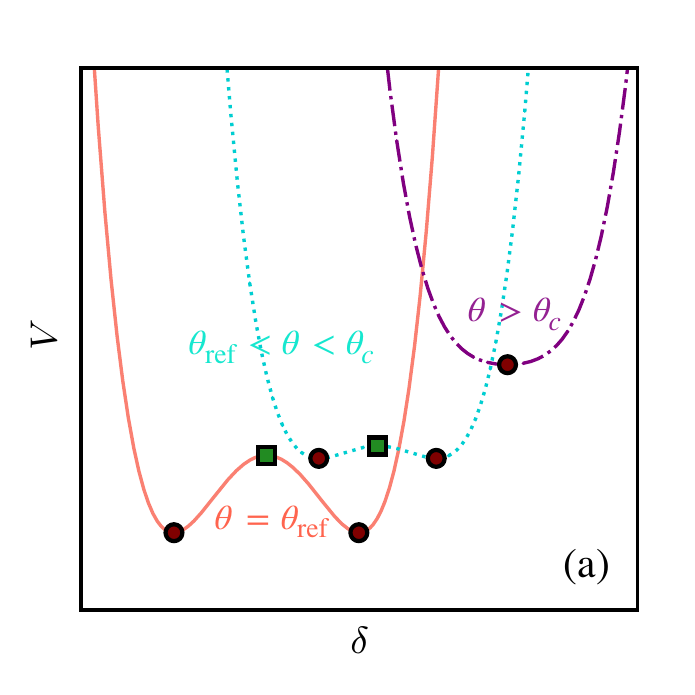} \centering} \subfigure[\label{fig:fig_1b}]
{\includegraphics[width=0.45\textwidth]{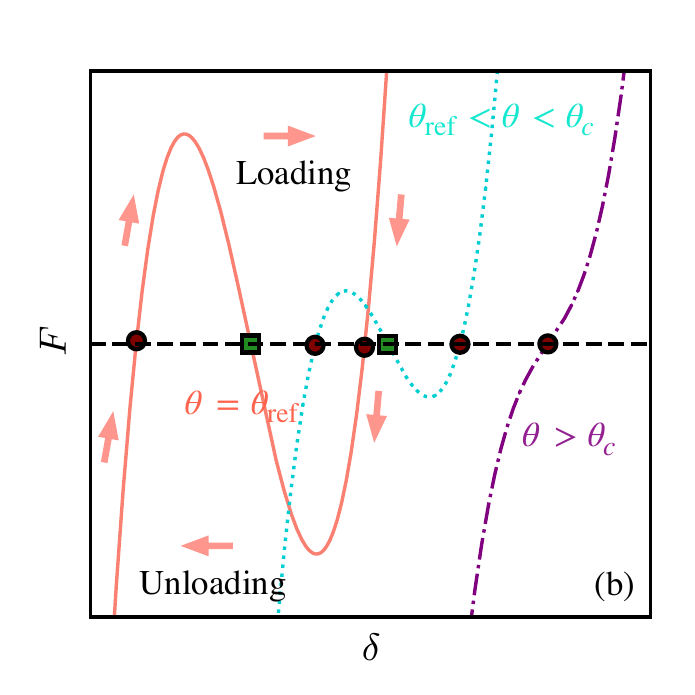} \centering}
\caption{\label{fig:fig_1} Representative energetic and equilibrium diagrams: (a) shows the
potential energy landscape while (b) shows the force-displacement curve for the proposed mechanical
system with increasing temperatures. Circles and squares correspond, respectively, to stable and
unstable points. At low temperatures (solid and dotted lines, $\theta < \theta_c$) the system has a
nonconvex potential energy with two stable and an unstable equibilibrium point and changes to convex
beyond a critical temperature $ \theta_c$ (dashdot line, $ \theta> \theta_c$) with a single stable
point. For a nonconvex potential, the force-displacement is non-monotonic and exhibits snap-through
instabilities that can be exploited for energy absorption as shown by arrows for $ \theta =
\theta_{\rm{ref}}$ in (b).}
\end{figure}

Instabilities emerge in solids and structures when their potential energy is not (quasi) convex
\citep{Knops1986}.  Interestingly, this property might depend on external factors to the system such
as forces and temperature. For example, consider a mechanical system whose potential energy is
depicted on the left of Fig.~\ref{fig:fig_1} for three different values of the
temperature~$\theta$. For $\theta = \theta_{\rm{ref}}$, this curve has two local minima and one
local maximum where the derivatives of the potential energy vanish, i.e., where the forces are zero
and the system is said to be \emph{in equilibrium}. The equilibrium point corresponding to the local
maximum is unstable since any small perturbation causes the system to jump to one of the local
minima. The force-displacement diagram for this system is shown on the right of Fig.~\ref{fig:fig_1}
and is non-monotonic for $\theta = \theta_{\rm{ref}}$, passing through zero at three points that
correspond to the two stable and one unstable equilibria. This lack of monotonicity is the fundamental
mechanism that is used in the design of \emph{bistable} systems where, beyond a certain load, the
structure snaps to another stable point as indicated by arrows in Fig. \ref{fig:fig_1}(b). Complete
\emph{reversibility} of the original state is achieved through an unloading cycle by applying a
force in the opposite direction as shown in Fig.~\ref{fig:fig_1}(b).

In this work, we present a theory for the design of bistable and reversible lattice
structures. Moreover, reversibility will be achieved through an external stimulus, temperature in
our case, that can be controlled at will and can be used to shape the potential energy. Similarly to
the system illustrated in Fig.~\ref{fig:fig_1}, the structures that will be proposed will possess
nonconvex potential energies at low temperatures that will become convex beyond a critical
temperature.  When the temperature is higher than the critical value, the force-displacement
relation will become monotonic and this will be exploited to drive the system back to a selected
minimum of the original configuration. These controlled transformations are inspired by the
thermomechanical cycle characteristic of shape-memory alloys such as Nitinol
\citep{Bhattacharya2004}, which exhibit multiple (energetically equivalent) variants of
martensite at low temperature and a single austenite configuration at high temperature. While
external loads can favor one particular martensite phase, the original configuration can be
recovered employing a heat treatment that transforms first the whole crystal to austenite and back
to the original martensite configuration on cooling. Our designed lattices thus mimic the
crystallographic phase-transformation mechanisms of shape-memory alloys at the structural lattice
scale by means of a judicious choice of geometry and thermomechanical properties.

This research finds resemblance with previous works that sought to design simple atomistic
models of shape memory alloys. For instance, \cite{pagano2003simple} developed a one-dimensional
model simulating shear in a two dimensional body described by a discrete system of masses
and nonlinear springs with mechanical behavior described by Lennard-Jones
potentials. \cite{pagano2003simple} established a link among the energies at different scales
(interatomic potential, the double-well potential which is a combination of interatomic potentials,
the equilibrium energy, the continuum energy), and identified lattice instabilities that were used
to explain the mechanics of solid phase transitions. Similarly, \cite{elliott2002stability}
developed an atomistic model to study the stability of thermally-induced martensitic transformations
in bi-atomic crystals using a discrete system of point masses with interactions described by Morse
potentials. Following \cite{pagano2003simple}, the fundamental hypothesis of
\cite{elliott2002stability} was that martensitic transformations are manifestations of lattice level
instabilities in certain crystals. Equilibrium solutions and their stability were examined as a
function of temperature to determine the crystal structures emerging from critical bifurcation
points. However, the authors pointed out that their model does not predict any temperature-induced
proper martensitic transformations because all the equilibrium paths with low symmetry which are
observed in shape-memory alloy martensites were are found to be unstable for the range of
temperatures investigated. Shortly after, \cite{kastner2003molecular,kastner2006molecular} developed
a two-dimensional molecular dynamics model for the investigation of crystalline austenite-martensite
phase transitions. The discrete model consisted of two types of mass points with interaction
functions of Lennard-Jones type which allowed to create stable square crystalline lattices which
transformed into sheared variants representative of the martensitic phases as a function of the
applied temperature. Unlike previous work of \cite{elliott2002stability}, the numerical simulations
of \cite{kastner2003molecular,kastner2006molecular} were shown to reproduce the fundamental features
of the martensitic transformation in shape memory alloys, e.g., the austenite being stable at high
temperature and the martensite at lower temperature, unloaded body transforms reversibly between
austenite and martensite under temperature control, etc. Additional molecular dynamics simulations
were performed by \cite{kastner2009mesoscale} to study post-transformation microstructure and moving
austenite–martensite interfaces. The calculations yielded to martensitic morphologies very similar
to real materials, including the nucleation of wedge-shaped, twinned martensite plates, plate growth
at narrow travelling transformation zones, etc. The molecular dynamics model of
\cite{kastner2003molecular,kastner2006molecular} was further applied by \cite{kastner2011molecular}
to study microstructure evolution during cyclic martensitic transformations. The cyclic loading was
shown to produce the accumulation of lattice defects so as to establish new microstructural elements
which represent a memory of the previous morphologies. These new elements were self-organised and
they provided a basis of the reversible shape memory effect in the model material. On the other
hand, \cite{hildebrand2008atomistic} carried out atomistic calculations using a two-dimensional
diatomic lattice --simplest system that exhibits a multi-well macroscopic potential-- to study the
nucleation and kinetics of shear induced detwinning in shape-memory alloys with interatomic
potential described by the Lennard-Jones. The calculations showed that the transformation rate is an
increasing function of shear stress and temperature, and that the transverse ledge propagation is
the mechanism underlying twin-boundary motion. Moreover, the effect of geometric nonlinearity on the
stability of two-dimensional mass-spring lattices was studied by \cite{friesecke2002validity}, who
derived the critical conditions for the mechanical behavior of the springs, and the critical strain
in the springs, for which the Cauchy-Born hypothesis fails. More recently,
\cite{elliott2006stability} investigated the equilibrium configurations and stability properties of
multi-atomic crystalline systems to determine the loading conditions leading to solid-to-solid
martensitic phase transformations. Stability criteria with respect to perturbations at the atomic
scale (phononstability) and at the continuum scale (homogenized-continuum-stability) were reviewed and
the so-called Cauchy-Born stability condition was introduced to provide an intermediate criterion by
considering perturbations at both the atomistic and continuum scales. \cite{elliott2006stability}
provided a unified presentation of these stability criteria for crystalline solids in equilibrium
configurations consistent with Cauchy-Born kinematics (uniform deformation and internal shifts of
sub-lattices).

In the past, innovative mechanical metamaterial designs have been proposed that exploit
thermomechanical coupling to generate interesting functionalities such as thermal
cloaking~\citep{Schittny2013}, structures with negative or low thermal expansion
coefficients~\citep{Lakes1996, Steeves2007, Wu2016, Boatti2017}, shape-morphing structures
\citep{Haldar2018}, or materials that can be actuated using shape memory polymers
\citep{Gdoutos2013, Guseinov2020}. Our main goal in this work is to guide the design of structural
lattices with bars of different materials so that the convexity of the potential energy can be
controlled with the temperature in a robust manner. To accomplish this objective rigorously, we
emphasize the analytical route and refer to other works where similar ideas have been explored
numerically (see, e.g., \cite{Klein2019}).

A complete structural analysis of bistable, reversible lattice materials demands tools from
nonlinear analysis that can be used to explore the entire design space, both qualitatively and
quantitatively. In particular, a rigorous stability analysis is mandatory to identify the main
parameters that affect the stability of the structure, their critical values when the stability
shifts, and a full characterization of the stability regions. In addition to these preliminary results, a
robust design requires the identification of \emph{all} the physical parameters that affect
qualitatively the nature of the stability, for example breaking the symmetry of the solution. For
this, we have chosen to employ \emph{singularity theory}~\citep{Golubitsky1985,Govaerts:00,Bazant:10}, a
branch of mathematics that provides a unified methodology to systematically study bifurcation
problems, revealing simultaneously \emph{all} possible perturbations of the governing equations that
qualitatively modify the stability of the system. Guided by this theory, the design space is
simplified to the maximum and numerical techniques can then be used to identify the numerical values
of the critical parameters. Remarkably, this approach guarantees that there remain no hidden
parameters that can affect the bifurcation behavior of the system.

Singularity theory provides a complete characterization of the stability and bifurcation diagram of a periodic lattice. Also, it reveals unambiguously the geometrical and/or material parameters that can be modified in order to change the shape of this diagram. This information is critical for the \emph{design} of bistable, reversible lattices, since it spares the effort to look for fruitless parameter combinations and reduces the exploration of the design space. After concluding this analysis, we have explored numerically selected examples involving \emph{finite} lattices that possess the bi-stability and reversibility features. These boundary value problems illustrate that the design is robust enough to carry on even when complete periodicity is loss and hint at potential applications of these lattices.

A summary of the remainder of this work is as follows. In Section~\ref{sec:2D-unit-cell} we use
singularity theory to analyze periodic two-dimensional lattices consisting of thermo-elastic bars,
identifying the parameters that guarantee the desired bi-stability and reversibility. One crucial
result of this part is the complete characterization of a \emph{phase diagram} for the structure,
one that can be later be used to select the geometry and material properties of the lattices.
Numerical examples of this lattice types will be shown in Section~\ref{sec:numerical-examples} that,
guided by the previous theoretical results, propose finite size lattices which exhibit the features
of bi-stability and reversibility. The analysis of Section~\ref{sec:2D-unit-cell} is extended to
three-dimensional lattices in Section~\ref{sec:3D-unit-cell}. These results are exploited in
Section~\ref{sec:3D-macroscopic} to carry out more complex three-dimensional examples of lattices
under full thermo-mechanical loading cycles and reveal some shape-memory-like
behavior. Section~\ref{sec:discussion} will close the article with a summary of the most relevant
results and some conclusions. For completeness and reference, the key concepts of singularity theory
are collected in \ref{sec:singularity-theory}.

\section{2D unit cell design of thermally reversible metamaterials}
\label{sec:2D-unit-cell}

\subsection{Description and kinematics}
In this section, we present a design for two-dimensional, bistable, lattice-based structures that are thermally
reversible and use singularity theory as a guide. We use a bottom-up approach and study first a
representative unit cell whose behavior, when arranged periodically, will result in a structure with
the desired thermomechanical properties.

\begin{figure}[t]%
  \centering%
  \includegraphics[height = 2.25 in]{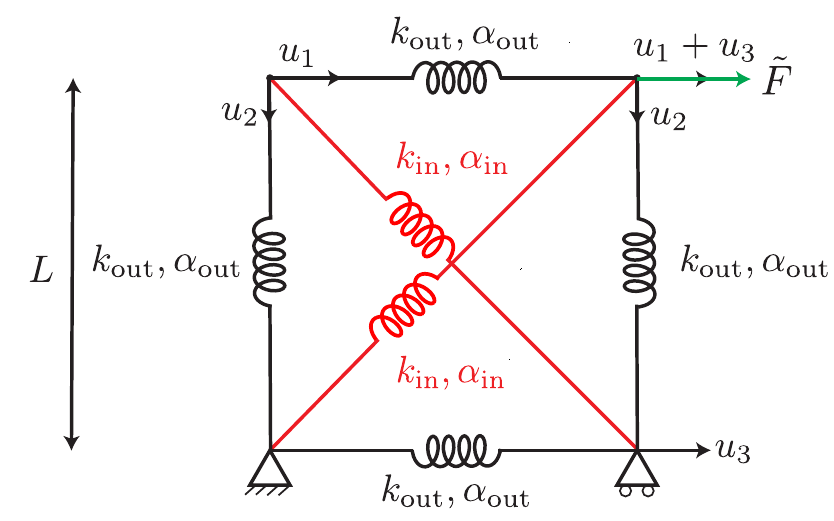}%
  \caption{Schematic of square 2D unit cell of length $L $ that has diagonal springs with
	stiffness and coefficient of linear thermal expansion $\kin, \alphain$ (shown in red) and an outer
	frame with properties $\kout, \alphaout$. The system is characterized by displacements $u_1, u_2$
	and $u_3$ and to measure the force-displacement response a force $\tilde{F}$ is applied on the top right
	corner.}%
  \label{fig:fig_2D_unitCell}%
\end{figure}

The unit cell consists of a deformable square frame of sides with length $L$ as depicted in
Fig.~\ref{fig:fig_2D_unitCell}. It is built with elastic springs whose thermoelastic response is
known. More specifically, the elastic constant and thermal expansion coefficient of
the springs in the outer frame are denoted as $\kout, \alphaout$, respectively, while the internal
springs, in turn, have elastic constant and thermal expansion coefficient $\kin,\alphain$,
respectively. Boundary conditions are applied on the two bottom vertices as illustrated in
Fig.~\ref{fig:fig_2D_unitCell}.

We study first the solution manifold for this structure and discuss the singularities that
might appear due to  temperature changes. Motivated by its application to the analysis of periodic
lattices, we analyze all possible affine deformations of the frame, namely, those of the form
\begin{equation}
  \mbs{x} = \mbs{F} \mbs{X} + \mbs{c}\ ,
  \label{eq:deformedCoor}
\end{equation}
where $\mbs{X}\in\mathbb{R}^2$ denotes the undeformed position of a point in the frame, $\mbs{x}$ is
its deformed position, and $\mbs{F},\mbs{c}$ are a constant tensor and vector, respectively. Imposing
boundary conditions that preclude any rigid body motion, the vector~$\mbs{c}$ must vanish and $\mbs{F}$
must be
\begin{equation}
  \label{eq-F}
  \mbs{F} =
   \begin{bmatrix}
    F_{11} & F_{12}  \\
   0 & F_{22} \\
  \end{bmatrix}\ ,
\end{equation}
for some constants $F_{11}, F_{12}, F_{22}$. Under these conditions, the motion of
the cell is determined by three displacements, $u_1, u_2$ and $u_3$
(cf. Fig.~\ref{fig:fig_2D_unitCell}) that satisfy
\begin{equation}
u_1 = F_{12} L\ , \qquad
u_2 =(F_{22}-1)L\ , \qquad
u_3 =(F_{11}-1)L\ .
\end{equation}

\subsection{Equilibrium equations}

To model the thermoelastic behavior of each spring we choose a free energy with expression
\begin{equation}
  \hat{V}(\lambda_i,\theta) =
  \frac{k_i}{2} \log^2 \frac{\lambda_i}{\lambda_i^0} - \alpha_i k_i \ (\theta - \theta_{\rm{ref}}) \log \frac{\lambda_i}{\lambda_i^0} + \varPsi(\theta),
  \qquad
  \varPsi(\theta) =
  c_0 \left( \theta - \theta_{\rm{ref}} - \theta \log \frac{\theta}{\theta_{\rm{ref}}}\right)\ .
 \label{eq:free-energy-hencky}
\end{equation}
where $c_0$ denotes the heat capacity, $\theta, \theta_{\rm{ref}}$ are the current and reference
temperatures of the structure, respectively, $\lambda_i, \lambda_i^0$ are the deformed and natural
lengths of the spring and $k_i, \alpha_i$ are its elastic constant and the coefficient of linear
thermal expansion and $\varPsi$ is a purely thermal contribution to the free energy.
We have chosen a specific form for the free energy function which is
proportional to the square of the logarithmic strain so that the elastic force softens for large
strains and also is non-symmetric for tension and compression. We note, however, that the
methodology and analysis shown below can be performed without loss of generality for any other form
of the free energy function (see, e.g., \cite{pagano2003simple}).

Expressing the deformed lengths $\lambda_i$ in Eq.~\eqref{eq:free-energy-hencky} as functions
of the displacements $\mbs{u}=(u_1,u_2,u_3)$, and adding the free energy of all the springs, the total free
energy of the system after a change in temperature $\Delta\theta$ reads
\begin{equation}
   \tilde{V}(\mbs{u},\Delta \theta) = \sum_{i} \hat{V}(\lambda_i(\mbs{u}),\theta_{\rm{ref}} + \Delta \theta)\ ,
\end{equation}
where $\Delta \theta = \theta - \theta_{\rm{ref}}$ is the temperature increase with respect
to the reference temperature. Defining the normalized temperature $\Theta = \alphaout \Delta\theta$,
a normalized free energy $\mcl{V}$ can be introduced via the relation
\begin{equation}
  \tilde{V}(\mbs{u},\Delta \theta) = \kout \mcl{V}(\mbs{u}, \Theta; ~\frac{\kin}{\kout},
\frac{\alphain}{\alphaout}) \ .
    \label{Eq:free-energy-normalized}
\end{equation}

The equilibrium equations for the frame are obtained by taking the derivatives of the free energy with respect to
the displacements and, following the notation of \ref{sec:singularity-theory}, they read
\begin{align}
  \mbs{g}(\mbs{u},\Theta; \frac{\kin}{\kout}, \frac{\alphain}{\alphaout})
  = \pd{\mcl{V}}{\mbs{u}}(\mbs{u}, \Theta; \dfrac{\kin}{\kout}, \dfrac{\alphain}{\alphaout}) = \mbs{0}\ .
   \label{Eq:eq-equations-2D}
\end{align}
This is a system of three nonlinear equations with three unknowns that depend on a control parameter, the nondimensional temperature $\Theta$, and two physical parameters of the structure $\sfrac{\kin}{\kout}$ and $\sfrac{\alphain}{\alphaout}$.

\subsection{Bifurcation}

We investigate the solutions of Eq.~\eqref{Eq:eq-equations-2D} for some fixed values of the
parameters while the control variable, the temperature, is varied. In particular, we examine
situations where the number of solutions of Eq.~\eqref{Eq:eq-equations-2D} changes with
temperature. For this purpose, we use singularity theory. The basic parts of this theory are
summarized in \ref{sec:singularity-theory}, and we refer to the monograph by Golubitsky and
Schaeffer \citep{Golubitsky1985, Golubitsky1988} for full details and complete proofs of the results
presented. Note that similar analysis on bifurcation theory with application of buckling problems
can be found, for instance, in \cite{budiansky1974theory}.

Let us choose, for example, $\sfrac{\kin}{\kout} = 0.5$ and $\sfrac{\alphain}{\alphaout} =
1000$. Our strategy consists, first, in identifying singular points and then, if they exist, using
singularity theory to investigate potential bifurcations. We select the reference
temperature $\theta_{\rm{ref}}$ such that the system is stress-free in its initial configuration
and gradually reduce the temperature from $\Theta = \alphaout \Delta \theta = 0$.

To begin the analysis, we first construct the Jacobian, $\mcl{L} = D_1\mbs{g}$ which is the Hessian
of the free energy~\eqref{Eq:free-energy-normalized}, and use its spectrum to characterize the
stability of the equation using the classical Lagrange-Dirichlet criterion.  For our particular
example, at $\Theta = 0$,  i.e. at the stress-free configuration, all the eigenvalues of
$\mathcal{L}$ are positive and the structure is stable. As the temperature is reduced, we find at a
critical temperature that one of the eigenvalues of $\mcl{L}$ turns negative and the point where the
eigenvalue vanishes corresponds to the singular point.  Interpreting this process physically, as we
reduce the temperature, and due to the differences between the coefficient of thermal expansion of
the outer frame and the diagonal springs, thermal forces are induced which eventually lead to
non-trivial solutions. For this particular example, the point of singularity is at
$(u_1^s,u_2^s,u_3^s,\Theta^s) = (0.0,-0.118L,-0.118L,-0.000625)$.

Once the singular point is identified, we apply the Liapunov-Schmidt reduction
(cf. \ref{subsec:Liapunov-schmidt}).  This manipulation allows to obtain, from the system of
equations~\eqref{Eq:eq-equations-2D}, a scalar equation that characterizes the singularity. We will
then test this reduced equation for the recognition conditions (see \ref{subsec:recogprob}) to
resolve the nature of the bifurcation, if any.

To begin the reduction process, we will first make a translation of the coordinates to
$\hat{\mbs{u}} = \mbs{u} - \mbs{u}^s$, $\hat{\Theta} = \Theta - \Theta^s$, so that the point of
singularity is now at the origin $(\hat{\mbs{u}}, \hat{\Theta}) = (\mbs{0},0) $. The Jacobian
$\mcl{L}$ at $(\hat{\mbs{u}}, \hat{\Theta}) = (\mbs{0},0)$ is
\begin{equation}
    \mcl{L}(\mbs{0},0) = \begin{bmatrix}
        0 & 0 & 0\\
        0 & 3.214 & 0 \\
        0 & 0 & 3.214
    \end{bmatrix}\ .
    \label{Eq:L-matrix}
\end{equation}
Clearly, at the point of singularity, $\rm{rank}~\mcl{L} = 3-1 = 2 $ and we can follow the procedure
of \ref{subsec:Liapunov-schmidt} to reduce the system of
equations~\eqref{Eq:eq-equations-2D}. First, we must choose the vector spaces
$\mcl{M}$ and $\mcl{N}$ that additively split $\mathbb{R}^3$ into $\rm{ker}~\mcl{L} \oplus
\mcl{M}$ and $\mcl{N} \oplus \rm{range}~\mcl{L}$.
The choices for $\mcl{M} =\rm{range}~\mcl{L} $ and $\mcl{N} =
\rm{ker}~\mcl{L}$ are reasonable choices. Next, we define the maps $E
:\mathbb{R}^3\to\rm{range}~\mcl{L}$ and $\mcl{L}^{-1} : \rm{range}~\mcl{L}\to\mcl{M}$ as
\begin{equation}
    E = \begin{bmatrix}
        0 & 0 & 0\\
        0 & 1 & 0 \\
        0 & 0 & 1
\end{bmatrix} \hspace{25pt}\rm{and}\hspace{25pt}
    \mcl{L}^{-1} = \begin{bmatrix}
        0 & 0 & 0\\
        0 & 0.311 & 0 \\
        0 & 0 & 0.311
   \end{bmatrix}\ .
\end{equation}
Note that even if $\sfrac{\kin}{\kout}, \sfrac{\alphain}{\alphaout}$ are changed and provided that
singular points still exist, the structure of $\mcl{L}$ and the above choices do not change. From
elementary linear algebra, $\rm{ker}~\mcl{L} $ is spanned by the basis vector $ (1,0,0 )$ and
$\rm{range}~\mcl{L}$ is spanned by the basis vectors $(0,1,0)~\rm{and}~(0,0,1)$. Thus, for the
choice  $\mbs{v_0} \in \rm{ker}~\mcl{L} $ and $\mbs{v_0}^* \in (\rm{range}~\mcl{L})^{\perp}$
one can select, for example,
$\mbs{v_0} = (1,0,0)$ and $\mbs{v_0}^* = (1,0,0)$. This completes the
definition for the reduction process and we now test directly for the derivatives of
Eqs.~\eqref{Eq:LSeqa}-\eqref{Eq:LSeqe}. For the case  $\sfrac{\kin}{\kout} = 0.5$
and $\sfrac{\alphain}{\alphaout} = 1000$, these derivatives evaluate to
\begin{align}
   \begin{split}
         g_{,x}(\mbs{0},0) &= 0 \ ,\\
         g_{,xx}(\mbs{0},0) &= 0 \ , \\
         g_{,xxx}(\mbs{0},0) &= 1.818\ , \\
         g_{,\Theta}(\mbs{0},0) &= 0 \ ,\\
         g_{,\Theta x}(\mbs{0},0) &= 513.739 \ .\\
   \end{split}
   \label{Eq:numReducedDerivatives}
\end{align}

The recognition problem for the pitchfork bifurcation~(cf. Eq.\eqref{Eq:pitchfork-conditions}) is
satisfied by relations~\eqref{Eq:numReducedDerivatives}. Moreover, since $g_{,xxx}\, g_{,\Theta\,x}
> 0$, it can be further concluded that the singularity identified corresponds to an \emph{inverted}
pitchfork (see \ref{subsec:recogprob}).

\begin{figure}[ht!]%
  \centering%
  \includegraphics[width=0.7\textwidth]{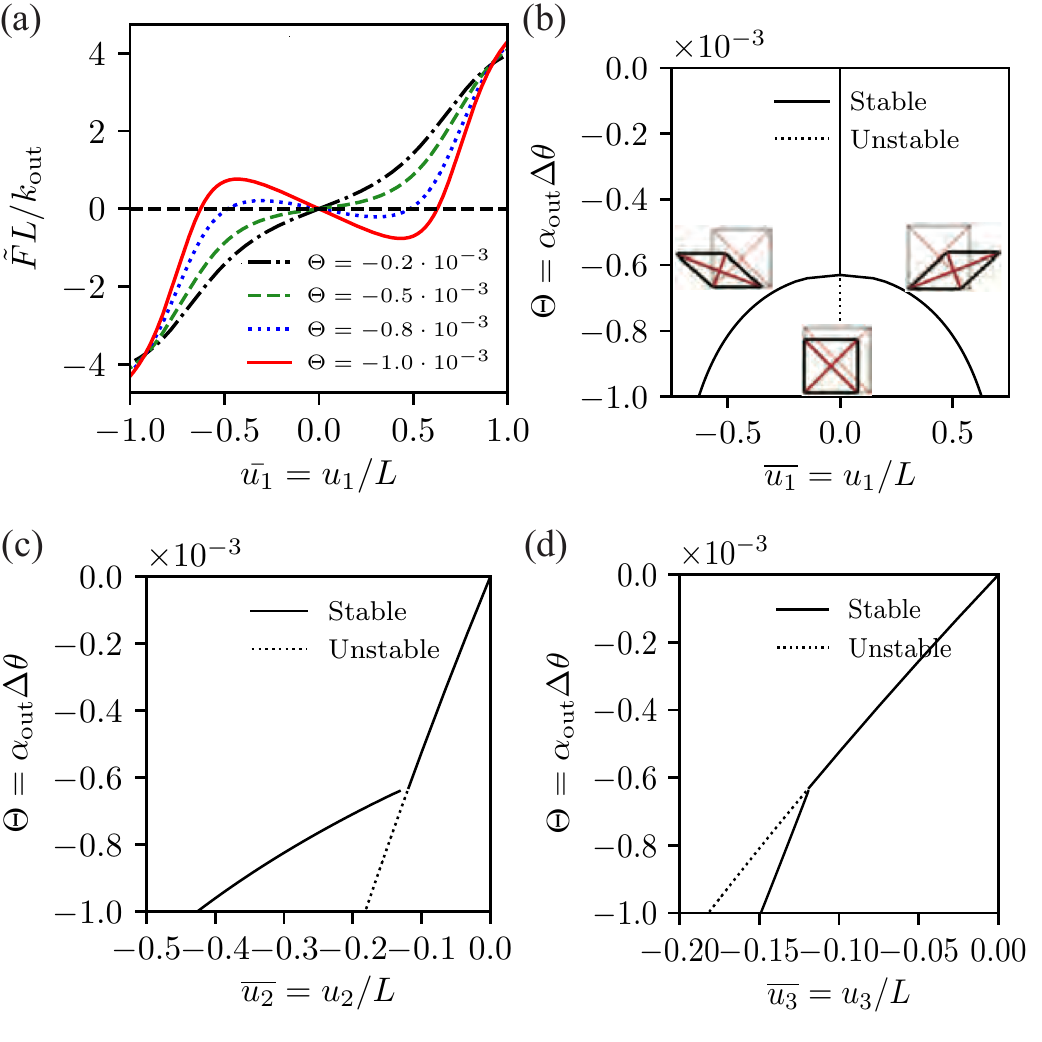}%
  \caption{Numerical solution of the 2D unit cell with $\sfrac{\kin}{\kout} =
	0.5$ and $\sfrac{\alphain}{\alphaout} = 1000$: (a) The force-displacement response that clearly shows a change in the monotonicity with 	change in temperature. (b)-(d) Bifurcation diagrams with temperature that exhibits a pitchfork bifurcation with respect to the horizontal displacement $u_1$ where multiple solutions arise beyond $ \Theta_s = -0.000625$. Shown in inset in (b) are the equilibrium shapes at $\Theta = -1 \cdot 10^{-3}$.}%
  \label{fig:2D_diagrams}%
\end{figure}

Let us now complement the information obtained from singularity theory with numerical solutions.
Let $\mbs{u}$ be the vector of displacements for the structure
and consider the application of an external force on the top right corner of the frame of value
$\mbs{\tilde{F}} = \Lambda \bar{\mbs{F}}$, where $\Lambda$ is a load scaling factor and $\bar{\mbs{F}}$ is a
unit point force in the horizontal direction (cf. Fig.~\ref{fig:fig_2D_unitCell}). Under these
conditions, the free energy of the frame is
\begin{equation}
  \mcl{\tilde{V}}(\mbs{u}, \Theta; ~\frac{\kin}{\kout}, \frac{\alphain}{\alphaout})
  = \kout  \mcl{V}(\mbs{u}, \Theta; ~\frac{\kin}{\kout},\frac{\alphain}{\alphaout}) - \Lambda u^R \ ,
      \label{Eq:free-energy-normalized-arclength}
\end{equation}
where $u^R = \bar{u_1} + \bar{u_3}$ is the horizontal displacement of the top right corner. Note
that the first row and column of the Jacobian~\eqref{Eq:L-matrix} are zero suggesting that non-trivial
solutions might appear in the~$u_1$ direction.

Given the lack of convexity of the free energy for arbitrary temperatures, the system is expected to
exhibit instabilities. Hence, if we employ a Newton-type solver to trace the equilibrium path of the
structure, we might encounter numerical problems when the tangent stiffness becomes singular. The
standard remedy for this situation consists in employing a path-following technique based on the
arc-length method \citep{Crisfield1981,Schweizerhof1986}. The general idea of this approach consists
in modifying the Newton solution, appending a new unknown, the generalized arc-length of the
solution, and a new equation to the global system of equilibrium equations. The arc-length $s$ is
defined over the generalized solution space, and its differential is
\begin{equation}
  \mathrm{d}s^2 = \mathrm{d}\Lambda^2 + |\mathrm{d} \mbs{u}|^2\ .
\label{eq:arclength}
\end{equation}
With this definition, at every step of the incremental solution for the structural equilibrium,
the following additional equation needs to be added
\begin{equation}
  f(\Delta \mbs{u}, \Delta\Lambda)
  = |\Delta \mbs{u}|^2  +
  \Delta \Lambda^2 \psi^2  - \Delta s^2 = 0\ .
  \label{eq:arclengthconstraint}
\end{equation}
Here, $\Delta \mbs{u} = \mbs{u} - \mbs{u}_{i-1}$ and $\Delta \Lambda = \Lambda -
\Lambda_{i-1}$, and the subscript $(i-1)$ refers to the last converged solution. A scaling
factor $\psi$ is introduced into the constraint~\eqref{eq:arclengthconstraint} to
render it dimensionally consistent. In practice, however, $\psi$ is often chosen to be zero (see, e.g.,
\citep{Bonet2008} for more details on the continuation technique).

Fig.~\ref{fig:2D_diagrams}(a) shows the equilibrium path for a design with
$\sfrac{\kin}{\kout} = 0.5$ and $\sfrac{\alphain}{\alphaout} = 1000$. It can be observed
in this figure that when the temperature is reduced,
the force-displacement response turns non-monotonic. Fig.~\ref{fig:2D_diagrams}(b) plots the
bifurcation diagram of the horizontal displacement, $u_1$, as a function of the normalized
temperature $\Theta$ which confirms the inverted pitchfork prediction by singularity theory (see
Section~\ref{subsec:recogprob}). Moreover,
equilibrium shapes of the cell at temperatures below the bifurcation point are shown
in the inset of Fig.~\ref{fig:2D_diagrams}(b). The solution at the center
is unstable and under any perturbation the cell snaps to either of the stable
configurations to the left or to the right.  Figs. \ref{fig:2D_diagrams}(c),(d) plot the bifurcation
diagram for the displacements $u_2, u_3$. These have only two branches below the bifurcation point
since, due to symmetry, the two stable branches in Fig.~\ref{fig:2D_diagrams}(b) coincide in
Figs. \ref{fig:2D_diagrams}(c) and (d).

\begin{figure}[ht!]%
  \centering%
  \includegraphics[width=0.85\textwidth]{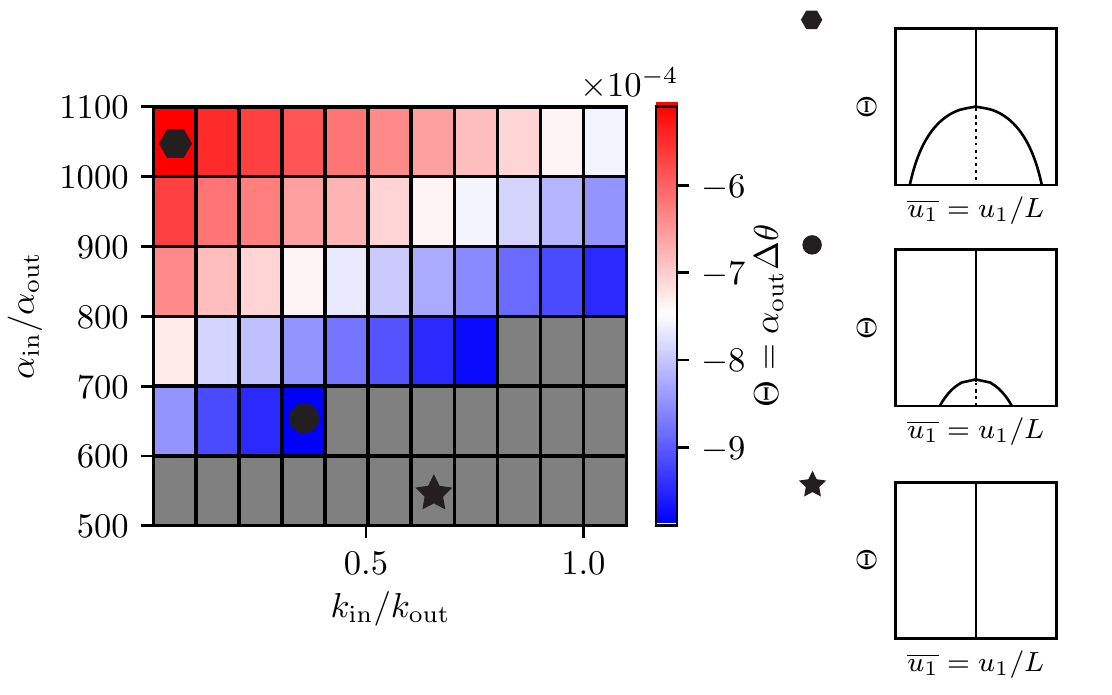}%
  \caption{Phase diagram of the unit cell for different structural parameters $\kin/\kout$ and
$\alphain/\alphaout$. Gray region corresponds to monostable behavior while colored regions represent
bifurcation behavior between a temperature range of $\Theta = 0$ to $\Theta = -1.0 \cdot 10^{-3}$
with the colormap showing the bifurcation temperature. The bifurcation diagrams corresponding to the
symbols in different regions of the phase-diagram are shown on the right. }%
  \label{fig:fig_7}%
\end{figure}

Finally, to study the influence of the ratios $\sfrac{\kin}{\kout}$ and
$\sfrac{\alphain}{\alphaout}$ we systematically vary these parameters and we search for bifurcations
between a range of normalized temperatures varying from $\Theta = 0$ to $\Theta = -1.0 \cdot
10^{-3}$. This corresponds to a change in a temperature of $100\ ^{\rm{o}}$C if $\alphaout \approx
10^{-5} ({^{\rm{o}}}C)^{-1}$. Fig.~\ref{fig:fig_7} depicts this phase diagram. In the figure, gray regions
correspond to points where no bifurcations are observed, while the colored regions represent
bifurcation behavior with the colormap indicating the bifurcation temperature. Note that as
$\sfrac{\kin}{\kout}$ decreases, the minimum value of the ratio $\sfrac{\alphain}{\alphaout}$ for
which a bifurcation occurs also decreases.  In fact, when $\kout\gg\kin$ the bistable behavior of
the structure only takes place for $\sfrac{\alphain}{\alphaout}> 600$ (see also
Section~\ref{Perturbation_limit_analysis}). A key outcome of the methodology developed in this
article is that it simplifies the identification of the range of values for $\sfrac{\kin}{\kout}$
and $\sfrac{\alphain}{\alphaout}$ for which the structural lattice becomes bistable and thermally
reversible.

\subsection{Perturbation analysis of the 2D unit cell}
\label{subsec:perturbation-analysis}

One of the central themes of singularity theory is the study of the effect of perturbations on the
solution set of an equation. In particular, the
universal unfolding of the pitchfork bifurcation is of the form
\begin{equation}
G(x,\lambda, \alpha_1, \alpha_2) = x^3 - \lambda x + \alpha_1 + \alpha_2 x^2\ .
\label{eq:univ-unfolding-pf}
\end{equation}
where $\alpha_1, \alpha_2$ are auxiliary parameters (see \ref{subsec:unfoldings}),
and thus it contains all the qualitatively possible perturbations of the pitchfork. Based
on this result, we proceed now to classify all the possible perturbations of the unit
cell, separating those that appear in the unfolding from those whose effects are qualitatively irrelevant.

\begin{figure}[ht!]%
  \centering%
  \includegraphics[width=\textwidth]{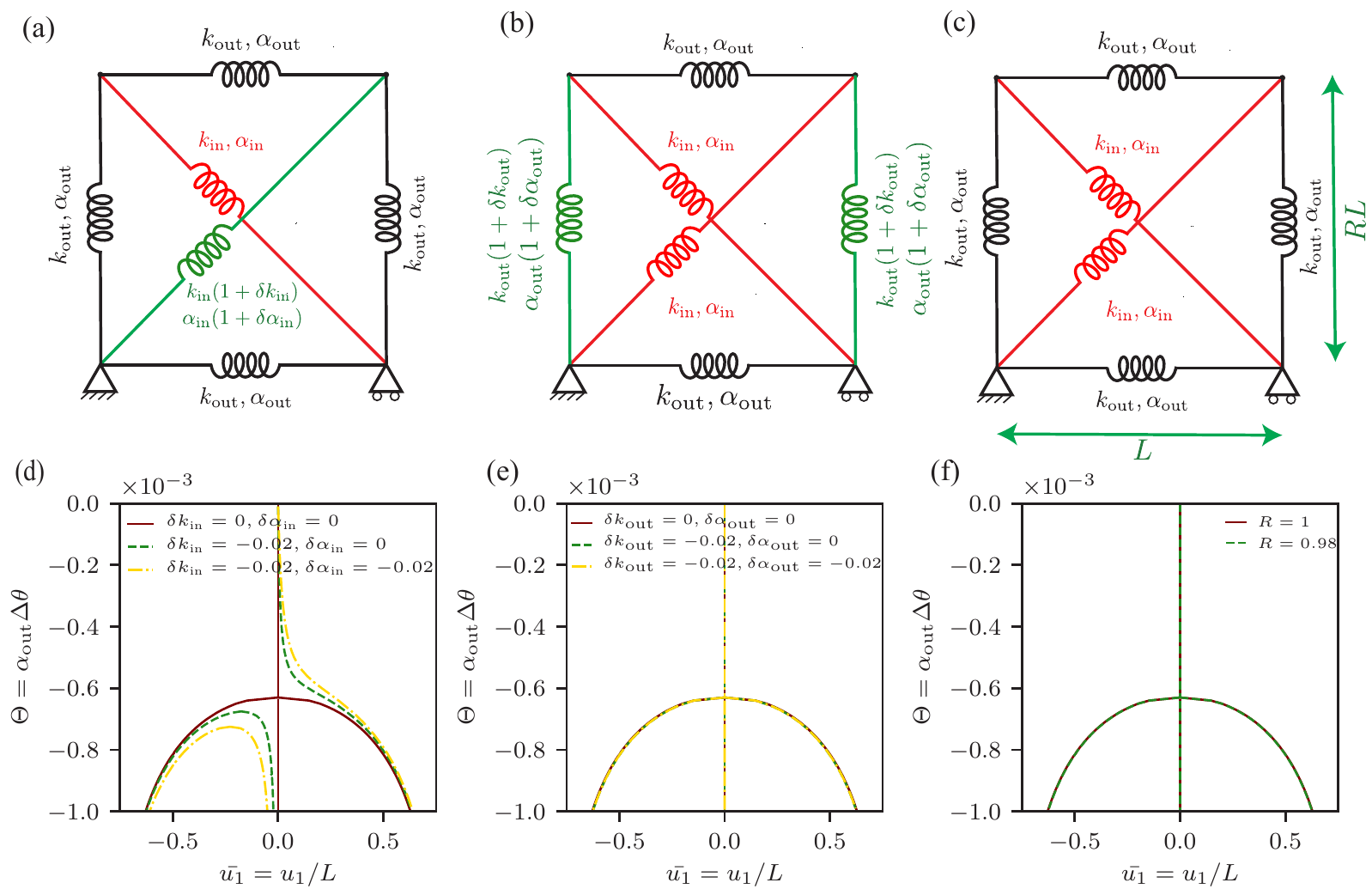}%
  \caption{Behavior of the structure under perturbation of some parameters of the unit cell. In (a)
the stiffness and the coefficient of thermal expansion of one of the diagonal springs is slightly
perturbed by $\delta \kin$ and $\delta \alphain$ (shown in green), respectively, while in (b) the
same properties but of a member in the outer frame are perturbed (also shown in green) by $\delta
\kout$ and $\delta \alphaout$, respectively. In (c) the aspect ratio of the unit cell is
modified. Moreover, (d), (e) and (f) show the bifurcation diagrams for (a), (b) and (c),
respectively. All the numerical examples use $\sfrac{\kin}{\kout} = 0.5$ and
$\sfrac{\alphain}{\alphaout} = 1000$.}%
  \label{fig:fig_8}%
\end{figure}

Considering again the periodic frame of Fig.~\ref{fig:fig_2D_unitCell}, we study the effect of
perturbations on the physical parameters of the system. First, let the stiffness and thermal
expansion coefficient of one of the diagonal springs be perturbed as in
$\kin\to \kin(1 + \delta\kin)$ and $\alphain\to\alphain(1+\delta\alphain)$, respectively, where
both $\delta \kin$ and $\delta \alphain$ are non-dimensional parameters (see
Fig.~\ref{fig:fig_8}(a)). In terms of these auxiliary parameters, the free energy of the system can be written as:
\begin{equation}
    \tilde{V}(\mbs{u}, \Delta \theta) = \kout \mathcal{V}_p (\mbs{u}, \Theta,  \delta \kin, \delta \alphain; ~\frac{\kin}{\kout}, \frac{\alphain}{\alphaout})\ .
\end{equation}
The equilibrium equation for the frame depends now on the auxiliary parameters as:
\begin{equation}
  \mbs{G}(\mbs{u},\Theta, \delta \kin, \delta \alphain; \frac{\kin}{\kout}, \frac{\alphain}{\alphaout})
  =
  \pd{\mcl{V}_p}{\mbs{u}}(\mbs{u}, \Theta,  \delta \kin, \delta \alphain; \dfrac{\kin}{\kout}, \dfrac{\alphain}{\alphaout})
  = \mbs{0}\ .
    \label{Eq:diagonal-pert-unfolding}
\end{equation}

The main goal now is to investigate the influence of the auxiliary parameters on the bifurcation
diagram. This can be tested by examining if the auxiliary parameters generate a universal unfolding
of the original bifurcation $\mbs{g}(\mbs{u},\Theta)$. To test for universal unfoldings, we must
first verify that under the absence of the auxiliary parameters, we recover the original equilibrium
equation $\mbs{g}$. Clearly, for $\delta \kin = \delta \alphain = 0$ this is satisfied, since
\begin{equation}
  \mbs{G}(\mbs{u}, \Theta, 0, 0; \frac{\kin}{\kout}, \frac{\alphain}{\alphaout})
  =\mbs{g} (\mbs{u}, \Theta; \frac{\kin}{\kout}, \frac{\alphain}{\alphaout})\ .
\end{equation}
Next, for the unfolding $\mbs{G}$ to perturb the
bifurcation diagram, for $\delta \kin, \delta \alphain \neq 0$, the recognition conditions of the
pitchfork (Eqs.~\eqref{Eq:pitchfork-conditions}), i.e. $ G = G_{,x} = G_{,\Theta} = G_{,xx} = 0$ and
$G_{,xxx}G_{,\Theta x} \neq 0$ (where $G$ is the reduced equation of $\mbs{G}$ after the
Liapunov-Schmidt reduction) must not be valid. If the recognition conditions are satisfied, then by
equivalence of Eq.~\eqref{Eq:equivalence}, $G$ can be transformed to the normal form of the
pitchfork, and the auxiliary parameters can be replaced by an equivalent single bifurcation
parameter, while the bifurcation diagram remains unperturbed. Indeed, for $\delta \kin, \delta
\alphain \neq 0$ and $\sfrac{\kin}{\kout} = 0.5, \sfrac{\alphain}{\alphaout} = 1000$, and using the
formulae of the Liapunov-Schmidt reduction (Eqs.~\eqref{Eq:LSeqa}-\eqref{Eq:LSeqe}), we find that
$G_{,x}\neq 0$ and $G_{,\Theta} \neq 0$, and hence the recognition conditions of the pitchfork
bifurcation are not verified.  Finally to determine if Eq.~\eqref{Eq:diagonal-pert-unfolding} is an
universal unfolding of $\mbs{g}$, we test for the recognition problem of the universal unfolding of
the pitchfork, i.e.,
\begin{align}
  \begin{split}
  \det \begin{bmatrix}
    0 & 0  & g_{,x\Theta} & g_{,xxx}\\
    0 & g_{,\Theta x}  & g_{,\Theta \Theta} & g_{,\Theta x x} \\
    G_{,\delta \kin} & G_{,\delta \kin x}  & G_{,\delta \kin \Theta} & G_{,\delta \kin x x} \\
    G_{,\delta \alphain} & G_{,\delta \alphain x}  & G_{,\delta \alphain \Theta} & G_{,\delta \alphain x x} \\
  \end{bmatrix}
   &=\det \begin{bmatrix}
    0 & 0  & 513.739 & 1.81813\\
    0 & 513.739  & 0 & 0 \\
    0.142 & 0.0321  & -255.01 & -0.164 \\
    0.177 & -0.161  & -319.06 & -90.5 \\
  \end{bmatrix} \\
   &= 3.38 \cdot 10^6 \neq 0\ ,
   \end{split}
\end{align}
where $g$ and $G$ are the reduced equations of $\mbs{g}$  and $\mbs{G}$, respectively, after the
Liapunov-Schmidt procedure and the derivatives are evaluated using
(Eqs.~\eqref{Eq:LSeqa}-\eqref{Eq:LSeqg}) at the point of singularity with $\delta \kin = \delta
\alphain = 0$. Since the condition for the recognition problem of the universal unfolding for the
pitchfork is met, $\mbs{G}(\mbs{u},\Theta, \delta \kin, \delta \alphain; \frac{\kin}{\kout},
\frac{\alphain}{\alphaout})$ is an universal unfolding of $\mbs{g}(\mbs{u},\Theta;
\frac{\kin}{\kout}, \frac{\alphain}{\alphaout})$ and it contains all the possible perturbations of
the pitchfork. For small changes in $\delta \kin $ and $\delta \alphain$, the bifurcation diagram is
expected to change drastically (see also Fig.~\ref{fig:fig_3}) as confirmed by
numerical solutions plotted in Fig.~\ref{fig:fig_8}(d).

Next, we perturb the stiffness and coefficient of thermal expansion of one of the members in the
outer frame (see Fig.~\ref{fig:fig_8}(b)) or the aspect ratio of the frame (see
Fig.~\ref{fig:fig_8}(c)). For both cases, we can construct the equilibrium equations $\mbs{G}$,
the unfolding of $\mbs{g}$, and verify that~$\mbs{G}$ satisfies the
recognition problem of the pitchfork bifurcation for non-zero perturbation. Thus, doing a change of
coordinates, the auxiliary parameters in this case can be completely replaced to reduce $G$ to the
normal form of the pitchfork bifurcation. We can conclude that these perturbations should not change
the bifurcation diagram at all as shown by their respective numerical solutions in
Figs.~\ref{fig:fig_8}(e) and~\ref{fig:fig_8}(f).

\begin{figure}[ht!]%
  \centering%
  \includegraphics[width=0.85\textwidth]{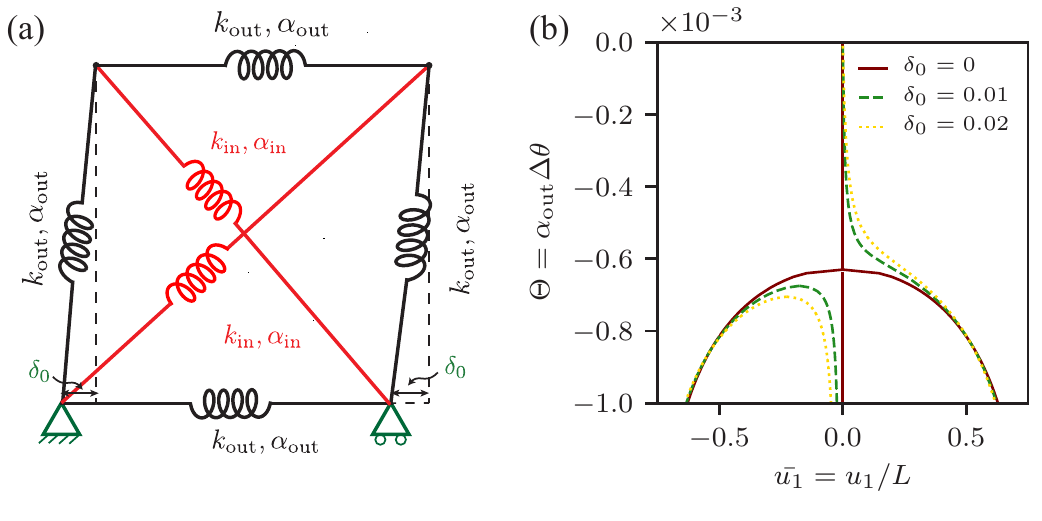}%
  \caption{A small defect $\delta_0$ is introduced at the support that also perturbs the bifurcation diagram as shown in (b).}%
  \label{fig:fig_9}%
\end{figure}

Using the idea of universal unfoldings of singularity theory, we have presented a novel strategy
where one can directly perform an analytical perturbation analysis to determine the main physical
parameters that can affect the bifurcation diagram of the structure. This is a key outcome of this
work and it helps in assessing the manufacturability of a lattice, enabling the identification of the
critical structural features where defects need to be avoided. In this structure, for example, a
defect in a diagonal spring will have a dramatic effect in the mechanical response of the lattice,
while defects in the springs of outer frame do not lead to qualitative differences in the
bifurcation diagram of the structure. Additionally, while we showed that perturbing the properties of
one of the diagonal springs generates a universal unfolding for the problem, it need not be the
only parameter that can affect the bifurcation diagram. For instance, a small defect at the support
can also perturb the stability of the lattice (see Fig.~\ref{fig:fig_9}). Singularity theory
shows, however, that this additional perturbation can not generate a qualitatively different
type of bifurcation from the ones already identified.

\subsubsection{Perturbation analysis in the limit $\kout \gg \kin$}
\label{Perturbation_limit_analysis}

Singularity theory proves that there are exactly are four
qualitatively different types of perturbations of the pitchfork bifurcation that depend on
$\alpha_1$ and $\alpha_2$ in the universal unfolding
(see \ref{subsec:unfoldings}). In Fig.~\ref{fig:fig_8}(a) we have shown that the auxiliary parameters
$\delta \kin$ and $\delta \alphain$ generate a universal unfolding of the unperturbed problem. In
practice, however, it is hard to relate the physical parameters $\delta \kin,  \delta
\alphain$ to the canonical parameters of the universal unfolding $\alpha_1, \alpha_2$ and thus
mostly result in the more commonly observable perturbation of the pitchfork
(Fig.~\ref{fig:fig_8}(d)).

In this section, we pay specific attention to the limit case $\kout \gg \kin$ anticipated in
Fig.~\ref{fig:fig_7} for which the value $\alphain/\alphaout$ triggering the bistable response of
the structure is minimum. In the limit $\kout \gg \kin$, i.~e. when the members of the outer frame
are almost rigid, the system can be simplified and a single nonlinear equation is required to
describe the equilibrium of the lattice. For $\kout \gg \kin$, we assume $u_3 \approx 0 $ and the
displacements $u_1$ and $u_2$ are related by
\begin{equation}
    u_1^2 + (1-u_2^2) = L^2\ .
  \end{equation}
In this case, the free energy of the system simplifies to
\begin{equation}
  \begin{split}
  \mcl{V}_{\rm{rigid}}(u_1,\Theta)
    =&
       \frac{\kin}{2} \left[ {\rm{log}}^2 \sqrt{1 - \frac{u_1}{L}}
       + {\rm{log}}^2 \sqrt{1 + \frac{u_1}{L}} ~\right]- \kin \Theta \left[\log \sqrt{1 - \frac{u_1}{L}} + \log \sqrt{1 + \frac{u_1}{L}}~\right]
\\
    &+ \varPsi(\theta_{ref}+ \alpha_{in}\Theta).
    \end{split}
\end{equation}
The equilibrium equation of the rigid frame is
\begin{equation}
    g_{\rm{rigid}}(u_1,\Theta) = - \pd{\mcl{V}_{\rm{rigid}}}{u_1}(u_1,\Theta) \ .
    \label{Eq:equilibrium-rigid-frame}
\end{equation}

\begin{figure}[t]%
\centering%
\includegraphics[width=\textwidth]{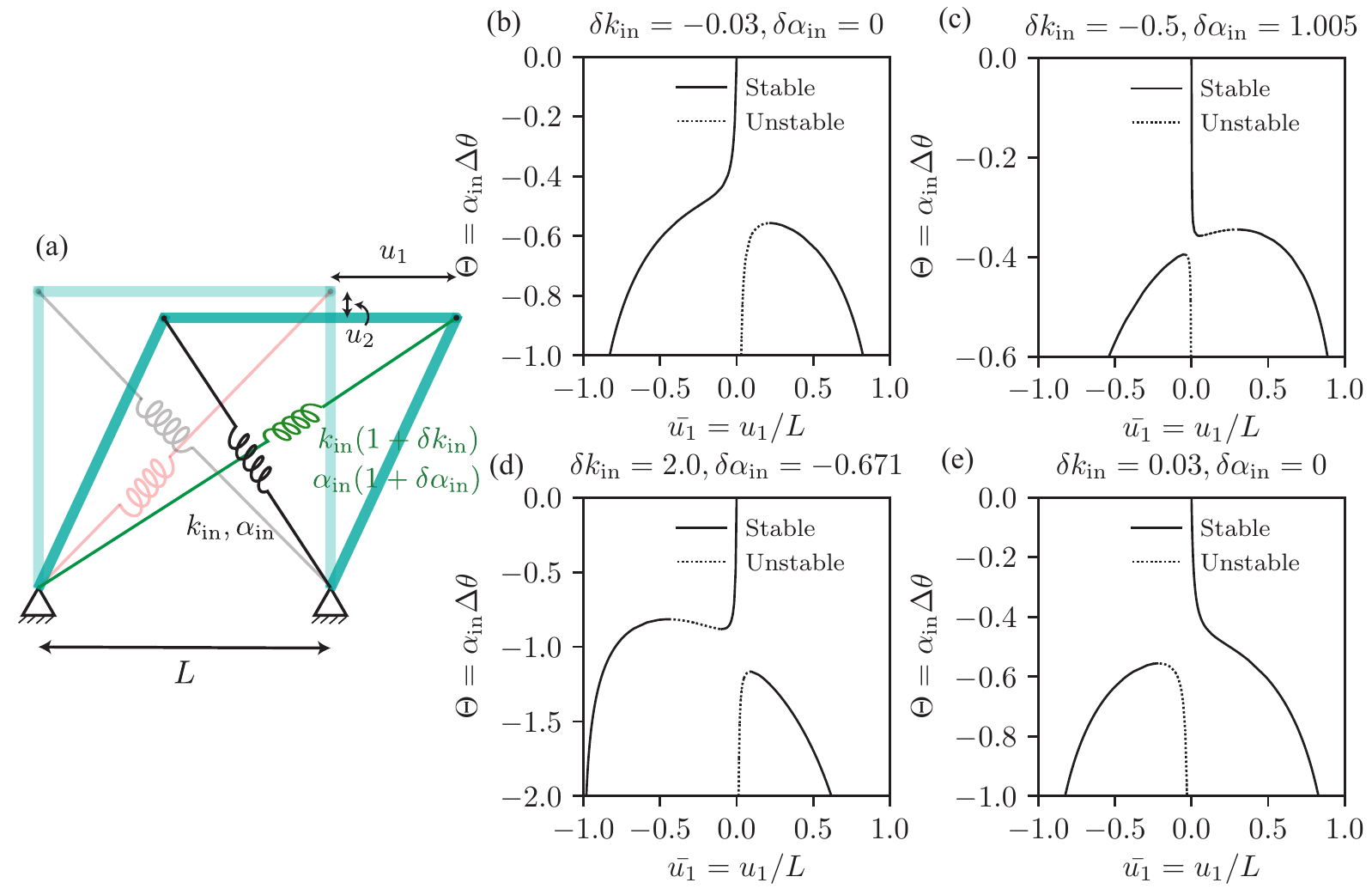}%
\caption{Perturbations of the pitchfork for a rigid frame: (a) shows the rigid frame with the diagonal springs
having slightly different stiffness and coefficient of linear thermal expansion by a factor $\delta \kin$ and $\delta \alphain $ respectively. Moreover, (b)-(e) are the bifurcation diagrams for different values of $\delta \kin $
and $\delta \alphain$. Bifurcation diagrams (c) and (d) are difficult to obtain without the
guidance of singularity theory.}%
 \label{fig:fig_10}%
\end{figure}

Eq.~\eqref{Eq:equilibrium-rigid-frame} has a singular point at $(u_1,\Theta) = (0,-0.5)$. Since
Eq.~\eqref{Eq:equilibrium-rigid-frame} is already a scalar equation, one can apply singularity
theory directly to verify that the nature of the bifurcation is indeed a pitchfork (see
\ref{subsec:recogprob}). In particular, when one of the diagonal springs is perturbed by $\delta
\kin$ and $\delta \alphain$, the equilibrium equation is modified to
\begin{equation}
  G_{\rm{rigid}}(u_1,\Theta,\delta \kin, \delta \alphain)
  = - \pd{\mcl{V}_{\rm{rigid}}}{u_1}(u_1,\Theta, \delta \kin, \delta \alphain) \ .
    \label{Eq:unfolding-rigid-frame}
\end{equation}
Approximating Eq.~\eqref{Eq:unfolding-rigid-frame} with a third-order Taylor expansion around the point
of singularity $(u_1,\Theta)=(0,-1/2)$ we get
\begin{align}
   \begin{split}
     G(u_1, \Theta, \delta \kin, \delta \alphain) \approx &
     \frac{1-(1+\delta \kin)(1+\delta\alphain)}{2}  -\frac{(1+\delta \kin)(\delta \alphain)}{2} \frac{u_1}{L} +
     (1+\delta \kin)\delta \alphain (\Theta + 1/2)  \\
     &+ \frac{(3\delta \kin + 2) - 2(1+\delta \kin)(1+\delta \alphain)}{2}\frac{1}{2}\left(\frac{u_1}{L}\right)^2  \\
     & + (1 + (1+\delta \kin)(1+\delta \alphain)) \frac{u_1}{L} (\Theta + 1/2)\\
     &+ 2(1+\delta \kin)\delta \alphain  (\Theta + 1/2) \left(\frac{u_1}{L}\right)^2\\
     &+ \frac{5 + 11(1+\delta \kin)-6(1+\delta \kin)(1+\delta \alphain)}{2} \frac{1}{6}\left(\frac{u_1}{L}\right)^3\ .
   \end{split}
   \label{eq:unfolding-taylor}
\end{align}
Comparing Eq.~\eqref{eq:unfolding-taylor} with the normal form of the universal unfolding of the
pitchfork (Eq.~\eqref{eq:univ-unfolding-pf}), we can relate the constants $\delta \kin$ and $\delta
\alphain$ with the parameters $\alpha_1$ and $\alpha_2$. This then allows us to probe the perturbed
bifurcation diagrams for different values of $\delta \kin$ and $\delta \alphain$ as shown in
Figs.~\ref{fig:fig_10}(b)-(e). Thus, when closed-form solutions are available, it is possible to
obtain the other two perturbations of the pitchfork that have the kinked shape shown in
Figs.~\ref{fig:fig_10}(c)-(d). For this, one must use specific values of $\delta \kin$ and $\delta
\alphain$ that are far from one and could not have been guessed, unless guided by singularity
theory.  This is another novel result of the methodology developed in this paper.


\section{Effective macroscopic behavior of 2D lattices}
\label{sec:numerical-examples}

\begin{figure}[h]%
    \centering%
    \includegraphics[width=\textwidth]{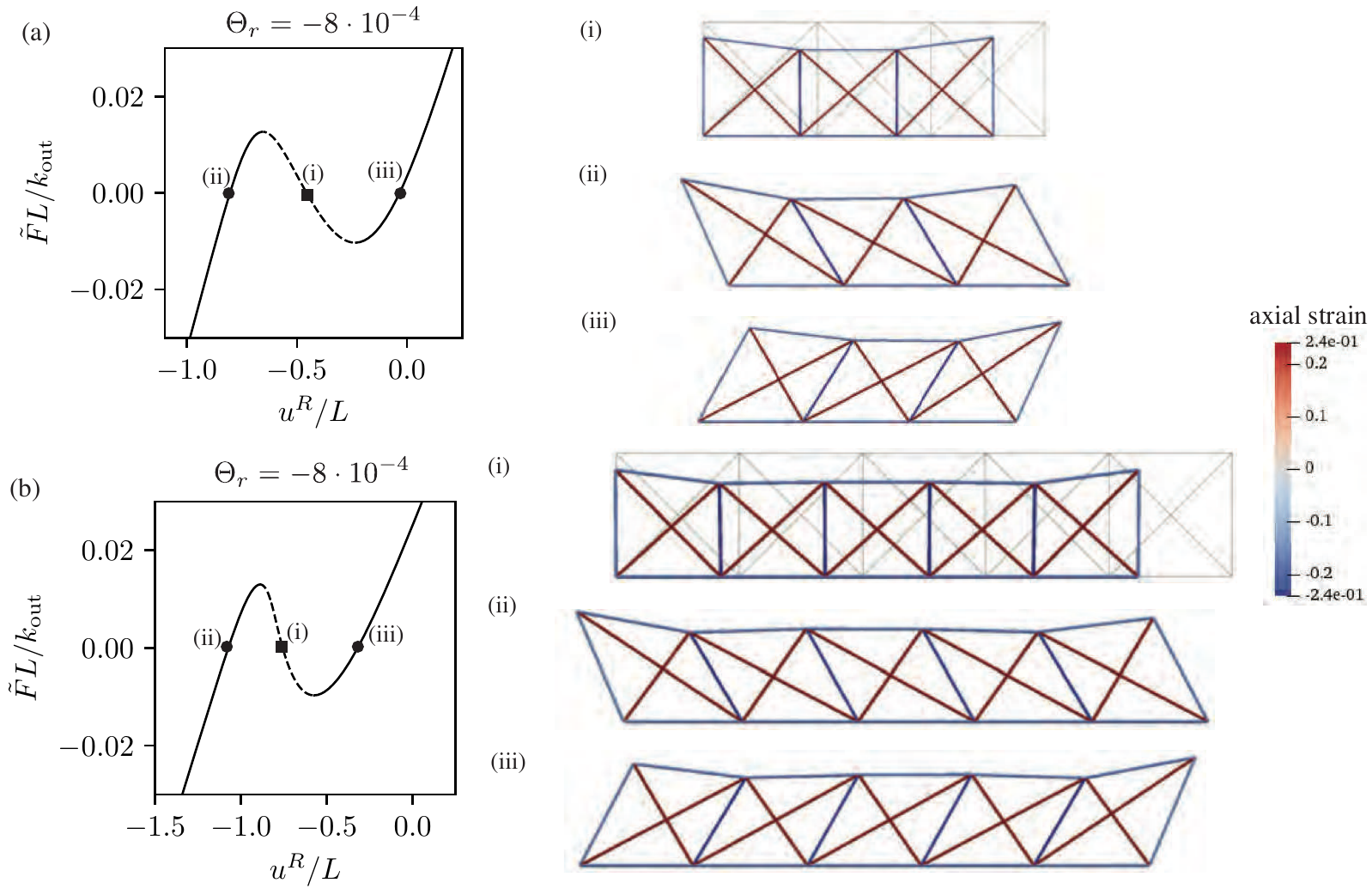}%
    \caption{1D array of $N$ unit cells: (a) and (b) show the force-displacement response for $N =
3$ and $N=5$, respectively, at $\Theta_r = -8\cdot 10^{-4}$.  The deformed configurations $(i)$,
$(ii)$ and $(iii)$ correspond to the equilibrium solutions in panels (a) and (b),
respectively. Shown in grey in (i) is the stress-free initial configuration at $\Theta = 0$.  The
parameters chosen for both panels are $\kin/\kout = 0.5$, $\alphain/\alphaout = 1000$.}%
\label{fig:fig-one-d-array}%
\end{figure}

\begin{figure}[ht!]%
\centering%
\includegraphics[width=\textwidth]{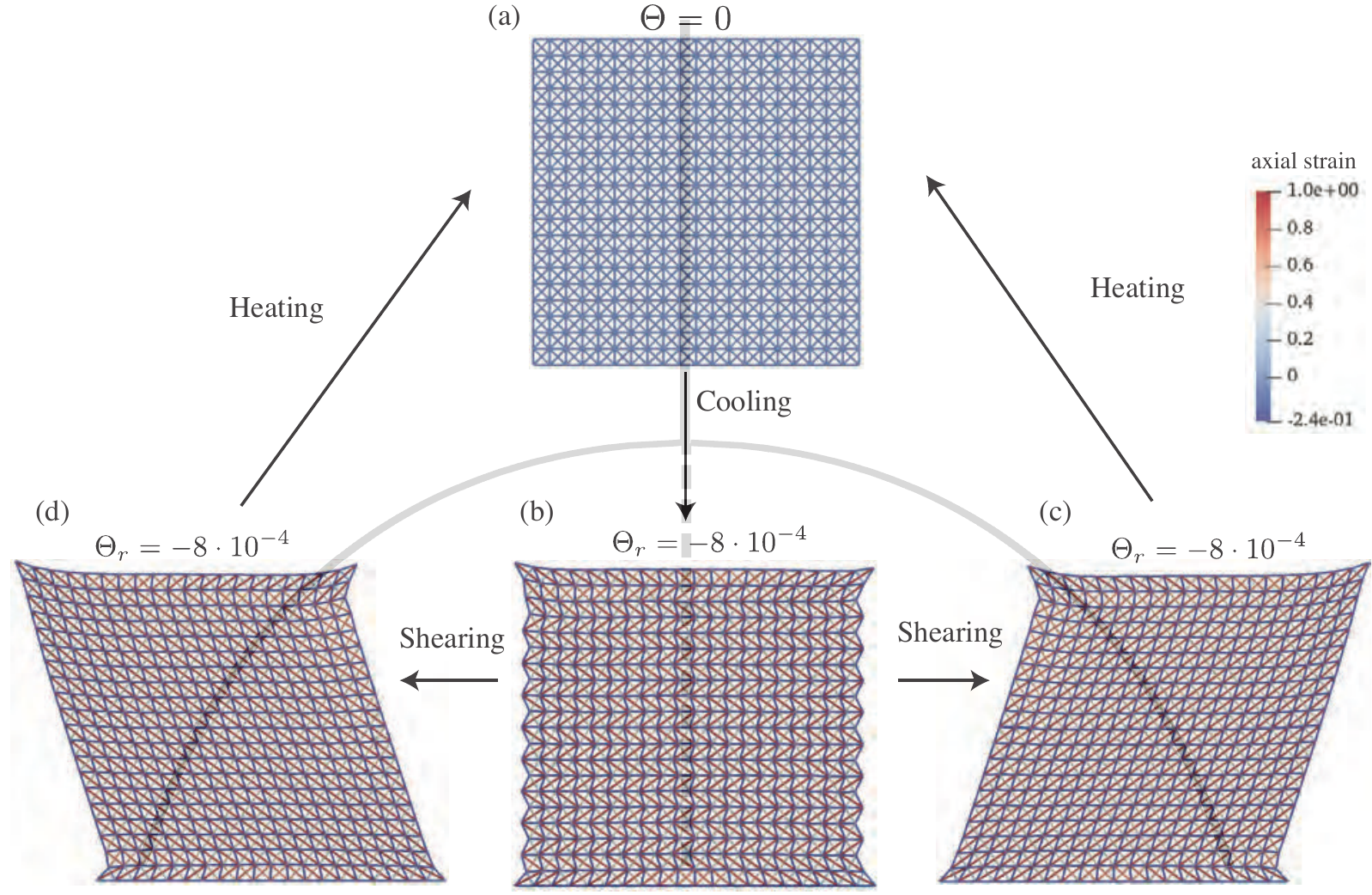}%
\caption{Shape memory behavior of the lattice: The stress-free configuration (a) is stable at
$\Theta=0$. When the temperature is reduced down to $\Theta = -8 \cdot 10^{-4}$, the structure
changes it shape to (b), where some cells buckle to the left and others to the right. Upon shearing
at ``cold'' temperature, the structure can deform from (b) to (c) or (d). Moreover, configuration
(a) can be completely recovered using a heat treatment, thus showing shape memory behavior. Data:
$20\times20$ unit cells and $\kin/\kout = 0.5$, $\alphain/\alphaout = 1000$. The colorbar of
the lattice structure is the axial strain in the bars which show both the diagonal members in tension while
the outer frame is in compression.}%
 \label{fig:complete-heat-cycle}%
\end{figure}

We would like to investigate next if lattices based on the cell designed in
Section~\ref{sec:2D-unit-cell} exhibit a behavior that is similar to the one of the unit cell.
Infinite
\emph{periodic} lattices can undergo affine deformation such as the one described in
Eq.~\eqref{eq:deformedCoor} and therefore certainly share the unit cell
behavior. We are interested here, however, in \emph{finite} lattices that can exhibit complex
motions and whose boundary effects might interfere with the stability properties.

Guided by the analysis of the unit cell, we now build macroscopic lattices stacking unit cells and
we check using simulations if the response of the unit cell is inherited at
the macroscopic scale. We begin by analyzing a one-dimensional array of $N$ unit cells where the
nodes at the bottom are constrained to move in the horizontal direction and the bottom-left corner is
completely fixed. We choose parameters $\kin/\kout = 0.5$ and $\alphain/\alphaout = 1000$ (the same
values used in Fig. \ref{fig:fig_8}) for which the unit cell exhibits a pitchfork bifurcation. We
begin our analysis by ramping down the temperature from $\Theta = 0$ to $\Theta_r=-8\cdot 10^{-4}$
(which is below the bifurcation temperature) employing a quasi-static simulation with Newton's
method. For two lattice structures, $N = 3$ and $N = 5$, the deformed structures are shown in
Fig.~\ref{fig:fig-one-d-array}(a)(i) and Fig.~\ref{fig:fig-one-d-array}(b)(i), respectively. Then we
measure the force-displacement response by applying a point force, $ \tilde{F}$ on the nodes on the
top surface using the arc-length method as detailed in Section \ref{sec:2D-unit-cell}. These curves
are shown in Figs.~\ref{fig:fig-one-d-array}(a)-(b), respectively, where $(i), (ii)$ and $(iii)$ are
the equilibrium positions. State $(i)$ represents an unstable configuration that can snap to either
$(ii)$ or $(iii)$ under small perturbations.

Next, we analyze a two-dimensional stacking of $N_x\times N_y$ unit cells. To study the whole
thermomechanical problem, we use a dynamic relaxation technique designed to simulate structures with
complex nonlinearities and instabilities. In this method, the static solution is obtained by
determining the steady-state response to the transient dynamic analysis of an ancillary
system. Since the transient solution is not necessary, fictitious mass and damping terms are added
to accelerate the convergence to the stationary solution (see \citep{Oakley1995a,Oakley1995b}
for details on the fundamentals and implementation of the method).

Fig.~\ref{fig:complete-heat-cycle} shows the complete thermomechanical cycle of a lattice.
In its initial state ($\Theta = 0 $), the structure is in a stress-free configuration
corresponding to Fig.~\ref{fig:complete-heat-cycle}(a). When the structure is cooled down to $\Theta
= -8\cdot 10^{-4}$, the symmetric configuration looses its stability, as predicted by the unit cell
analysis. As in any numerical computation, round-off errors trigger random perturbation of the
equilibrium and hence the cells snap to either the left- or right-tilted (stable) configurations, as
depicted in Fig.~\ref{fig:complete-heat-cycle}(b). If, still at the same temperature, shearing
forces are applied on the lattice, it can deform to either configuration (c) or (d). The colorbar in
(b), (c) and (d) is the axial strain and it is interesting to note that both the diagonal strings
are in tension while the outer members are in compression which is slightly
counter-intuitive. Finally, if the temperature of the structure is raised to its initial value
($\Theta=0$), the symmetric configuration becomes the only stable one and the system returns to its
initial position. Note that all the steps of the thermomechanical cycle of the structure were
predicted by the singularity theory, as illustrated by the \textit{shaded} inverted pitchfork included
in the graph behind the contours obtained from the finite element calculations.

This cycle shows that, with the chosen parameters, the behavior of the lattice mimics that of
a shape memory material. By tailoring the response of the unit cell, a macroscopic behavior
is obtained which exhibits a single stable phase at ``high'' temperature, and several
stable phases at ``low'' temperature that are energetically equivalent. Remarkably,
and similarly to true shape-memory materials, the thermomechanical cycle is completely reversible.



\section{3D unit cell design of thermally reversible metamaterials}
\label{sec:3D-unit-cell}
\begin{figure}[ht!]%
\centering%
\includegraphics[width=\textwidth]{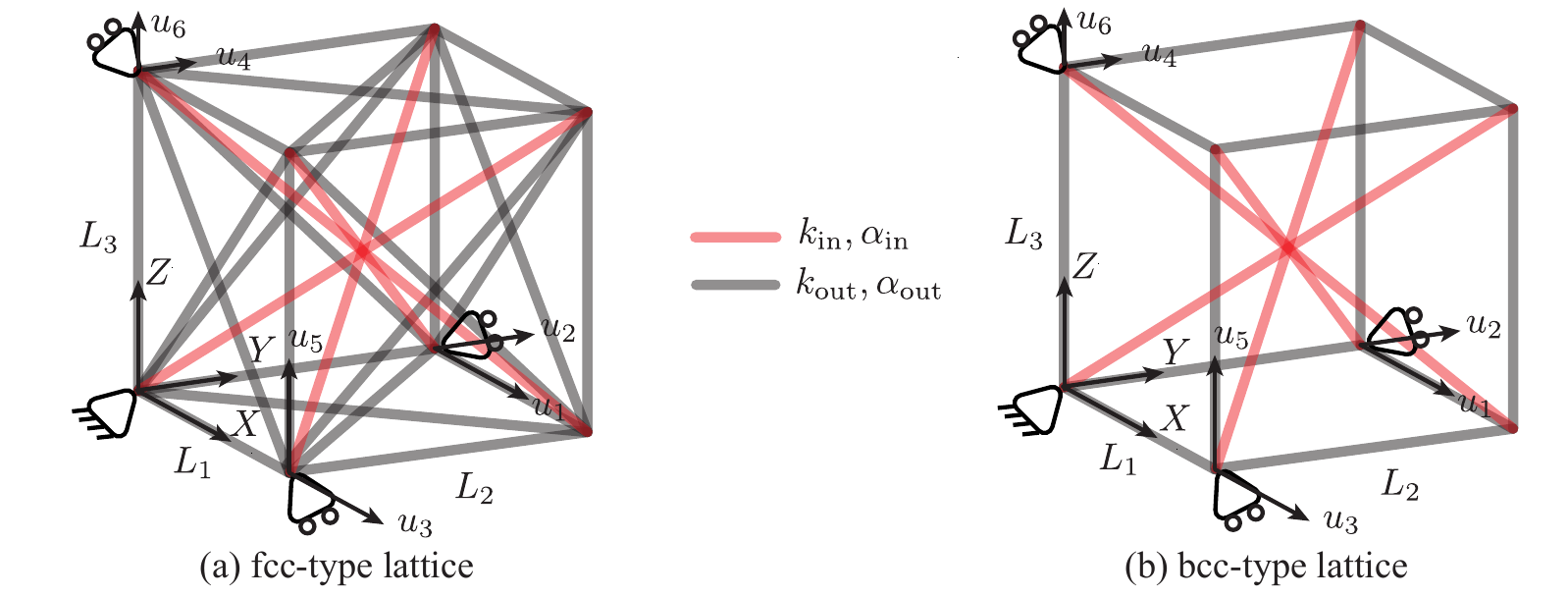}%
\caption{3D orthorhombic unit cells where (a) is for convenience referred to as an fcc type lattice
that has bars along the edges and the face diagonals with properties $\kout, \alphaout$ and body
diagonals with properties $ \kin, \alphain$ while (b) is a bcc type lattice that only has bars along
the edges and the body diagonals with properties ($\kout, \alphaout$), ($\kin, \alphain$)
respectively.}%
 \label{fig:3d-unit-cell}%
\end{figure}

A natural extension of the two-dimensional lattices studied in Section~\ref{sec:numerical-examples} to
three-dimensional geometries can be pursued by considering orthorhombic lattices with a rectangular
base ($L_1$ by $L_2$) and height ($L_3$) as shown in Fig.~\ref{fig:3d-unit-cell}.  We study
two types of such lattices: (a) fcc-type lattices where edges and face diagonals are bars with
stiffness and coefficient of thermal expansion $\kout, \alphaout$ while body diagonals have
properties $\kin, \alphain$ respectively; (b) bcc-type lattices that are the same as in (a), but
without bars on the face diagonals.

Similarly to the analysis of Section~\ref{subsec:perturbation-analysis}, we consider all possible affine
deformations of the frame with expression
\begin{equation}
\mbs{x} = \mbs{F}\mbs{X} + \mbs{c}\ ,
\end{equation}
where $\mbs{X}\in\mathbb{R}^3$ denotes the undeformed position of a point in the frame, $\mbs{x}$ is
its deformed position, and $\mbs{F},\mbs{c}$ are a constant tensor and vector, respectively. Boundary conditions
are imposed as shown in Figs.~\ref{fig:3d-unit-cell}(a),(b). Here one vertex is fixed and three vertices on the
$X, Y, Z$ axes are constrained on the $XZ, XY, YZ$ planes, respectively, to avoid rigid body rotations.
Under these conditions, the tensor $\mbs{F}$ must be of the form:
\begin{equation}
\mbs{F} = \begin{bmatrix}
		F_{11} & F_{12} & 0\\
		0 & F_{22} & F_{23} \\
		F_{31} & 0 & F_{33}\\
		\end{bmatrix}\ .
\end{equation}
The motion of the unit cell is thus uniquely determined by six displacements
\begin{equation}
   \begin{split}
        u_1 &= F_{12}L_2\ , \\
        u_2 &= (F_{22}-1)L_2\ , \\
        u_3 &= (F_{11}-1)L_1\ ,
  \end{split}
   \quad \quad  \quad \quad
  \begin{split}
       u_4 & = F_{23}L_3\ , \\
       u_5 &= F_{31}L_1\ ,\\
       u_6 &= (F_{33}-1)L_3\ ,
   \end{split}
 \end{equation}
 that can be related to the displacements of the three vertices on the $X, Y, $ and $Z$ axes.
 By adding the free energy of all the bars, we derive the
total free energy:
\begin{equation}
   \kout \mcl{V}(\mbs{u}, \Theta; ~\frac{\kin}{\kout}, \frac{\alphain}{\alphaout}) = \sum_{i} \hat{V}(\lambda_i(\mbs{u}),\theta_{\rm{ref}} + \Delta \theta)\ ,
\end{equation}
where $\mbs{u} = (u_1, u_2, u_3, u_4, u_5, u_6)$, $\Theta = \alphaout\, \Delta \theta$ is the
normalized temperature and $\hat{V}(\lambda, \theta)$ is the free energy of the each of the
individual springs (see Eq.~\eqref{eq:free-energy-hencky}). The equilibrium equations then follow from
the derivative of the free energy as
\begin{align}
  \mbs{g}(\mbs{u},\Theta; \frac{\kin}{\kout}, \frac{\alphain}{\alphaout})
  \equiv
  \pd{\mcl{V}}{\mbs{u}}(\mbs{u}, \Theta; \dfrac{\kin}{\kout}, \dfrac{\alphain}{\alphaout}) = \mbs{0}\ ,
   \label{Eq:eq-equations-3D}
\end{align}
which is analogous to Eq.~\eqref{Eq:eq-equations-2D}, now for the three-dimensional cell.
Since the kinematics of the unit cell is determined by the solution of six coupled nonlinear equations, the analysis
becomes more complex than in the plane case.

We will study next four types of three-dimensional lattices.

\subsection{Orthorhombic fcc lattices ($ L_1 \neq L_2 \neq L_3$)}
\label{subsec:orthofcc}
We break the symmetry in fcc-type lattices by choosing all sides of the unit cell to be of different
length. This makes the analytical derivations significantly simpler than in the symmetric case.
Specifically, we choose $\dfrac{L_2}{L_1} = 2$, $\dfrac{L_3}{L_1} = 3$, $\dfrac{\kin}{\kout} = 1.0$
and $\dfrac{\alphain}{\alphaout} = 1000$. To investigate possible instabilities in the system we
look at the eigenvalues of the Jacobian $\mcl{L}(\mbs{u}, \Theta) = D_1\mbs{g}(\mbs{u}, \Theta)$ as
we vary the normalized temperature $\Theta$. At $\Theta = 0$, the stress-free configuration, all the
eigenvalues are positive. When $\Theta=\Theta_1^s=-0.00188$, one of the eigenvalues of the Jacobian
vanishes and we identify the first singular point $(\mbs{u}_1^s, \Theta_1^s)$. The Jacobian at the
point of singularity is
\begin{equation}
\mcl{L}(\mbs{u}_1^s, \Theta_1^s) = \begin{bmatrix}
0.3073 & 0 & 0 & 0 & 0 & 0 \\
0 & 3.649 & 0.52& 0 & 0 & 0 \\
0 & 0.52 & 5.795 & 0 & 0 & 0.213 \\
0 & 0 & 0 & 0 & 0 & 0 \\
0 & 0 & 0 & 0 & 0.47 & 0 \\
0 & 0 & 0.214 & 0 & 0 & 2.782 \\
\end{bmatrix}\ .
\label{eq:jacobian-3D-1}
\end{equation}
Since only one eigenvalue vanishes at the singular point, we can use the Liapunov-Schmidt reduction
as presented in~\ref{subsec:Liapunov-schmidt} and reduce the six degrees of freedom equation to a
scalar equation $g(x,\bar{\theta})$ whose derivatives can be directly computed using formulae
Eqs.~\eqref{Eq:LSeqa}-\eqref{Eq:LSeqg}. At the singular point $(\mbs{u}_1^s, \Theta_1^s) $ these
are:
\begin{equation}
   \begin{split}
       g(\mbs{u}_1^s, \Theta_1^s) &= 0\ , \\
       g_x(\mbs{u}_1^s, \Theta_1^s) &= 0\ , \\
       g_{xxx}(\mbs{u}_1^s, \Theta_1^s) &= 0.361\ ,
  \end{split}
   \quad \quad  \quad \quad
  \begin{split}
        g_{\bar{\theta}}(\mbs{u}_1^s, \Theta_1^s) &= 0\ , \\
        g_{xx}(\mbs{u}_1^s, \Theta_1^s) &= 0\ , \\
        g_{\bar{\theta} x}(\mbs{u}_1^s, \Theta_1^s) &= 197.12\ .
   \end{split}
   \label{eq:reg-cond-3D-1}
\end{equation}

These equations satisfy the recognition conditions for a pitchfork bifurcation (see
\ref{subsec:recogprob}) and it can be concluded that there will be
such a singularity at $(\mbs{u}_1^s, \Theta_1^s)$, with the bifurcation in displacement $u_4$. Note
that the fourth row and column of the Jacobian~\eqref{eq:jacobian-3D-1} vanish. Moreover
$g_{xxx}(\mbs{u}_1^s, \Theta_1^s)  g_{\bar{\theta} x}(\mbs{u}_1^s, \Theta_1^s) > 0$, hence the
bifurcation can be classified as an inverted pitchfork.

\begin{figure}[ht!]  \centering
\subfigure[\label{fig:fccOrthoa}]{\includegraphics[width=0.32\textwidth]{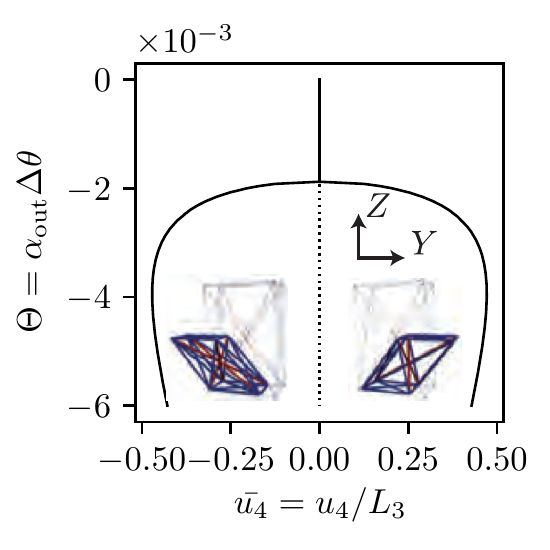} \centering}
\subfigure[\label{fig:fccOrthob}]{\includegraphics[width=0.32\textwidth]{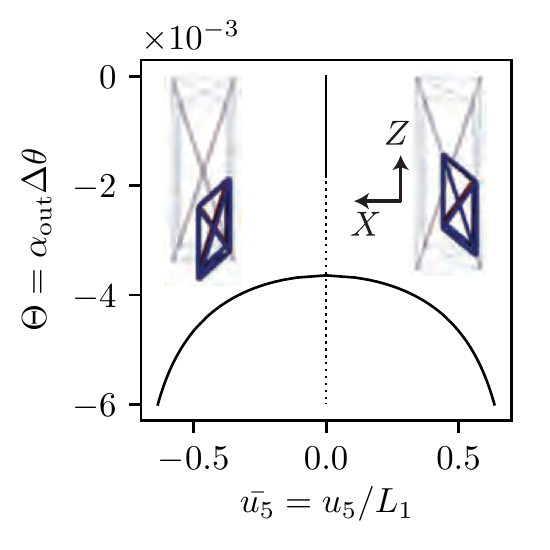} \centering}
\subfigure[\label{fig:fccOrthoc}]{\includegraphics[width=0.32\textwidth]{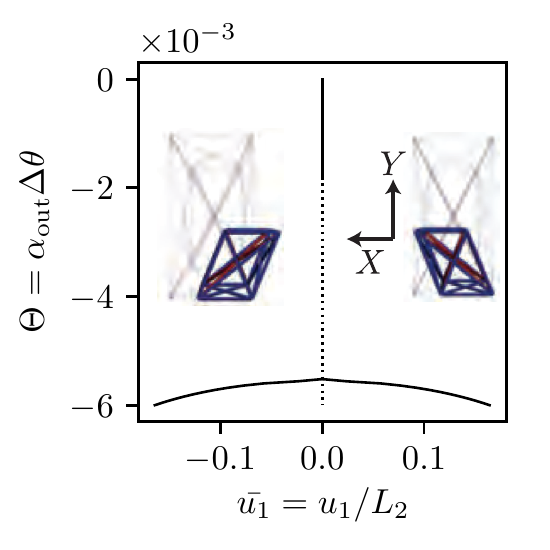} \centering}
\subfigure[\label{fig:fccOrthod}]{\includegraphics[width=0.32\textwidth]{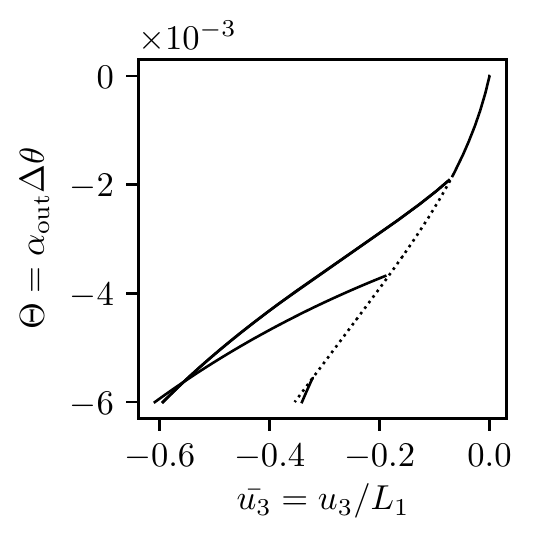} \centering}
\subfigure[\label{fig:fccOrthoe}]{\includegraphics[width=0.32\textwidth]{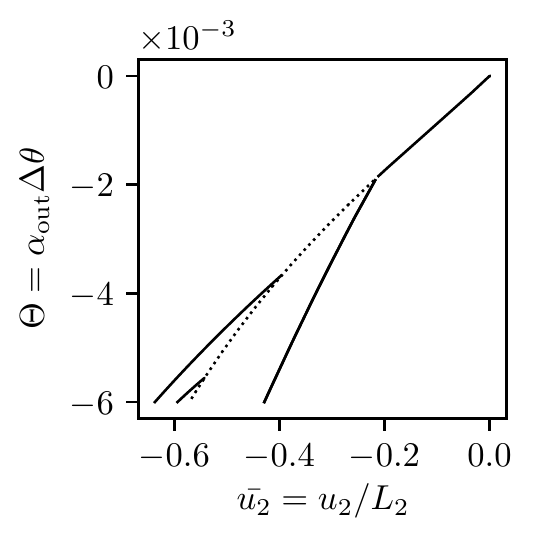} \centering}
\subfigure[\label{fig:fccOrthof}]{\includegraphics[width=0.32\textwidth]{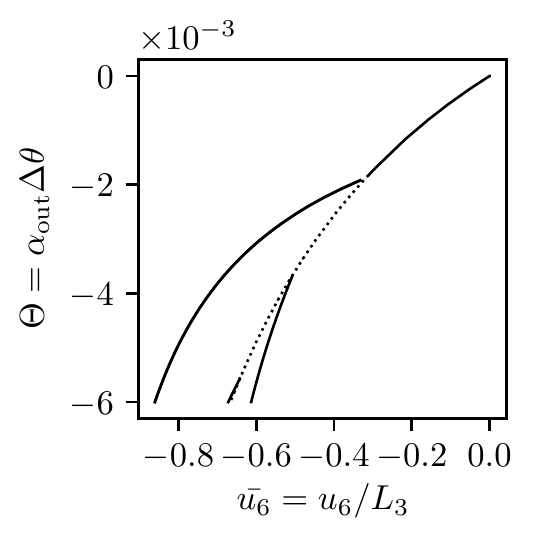} \centering}
\caption{\label{fig:fccOrthorhombic} Bifurcation diagrams for orthorhombic fcc type lattices with
$\dfrac{L_2}{L_1} = 2$, $\dfrac{L_3}{L_1} = 3$, $\dfrac{\kin}{\kout} = 1.0$,
$\dfrac{\alphain}{\alphaout} = 1000$. One clearly observes three different pitchforks at three
different temperatures in the three directions $Y, Z$ and $X$ respectively as shown by the
respective shear displacements in (a), (b) and (c). Figures (d), (e) and (f) plot the bifurcation
diagrams of the axial displacements. Insets in (a), (b) and (c) show the bifurcated lattice in its
stable branches of their respective pitchforks. Solid lines are stable solutions while dotted lines
are unstable solutions.}
\end{figure}

Next, we study the effect of reducing the temperature even further. We observe that a second and a
third eigenvalue of $\mcl{L}(\mbs{u}, \Theta)$ vanish at $\Theta_2^s = -0.00364$ and $\Theta_3^s =
-0.00555$, suggesting that there are two additional points of singularity at $(\mbs{u}_2^s,
\Theta_2^s)$ and $(\mbs{u}_3^s, \Theta_3^s)$, respectively. Following the same approach as for the
first point of singularity, we can again show that there are inverted pitchforks in the solution
at these two points with pitchforks in the direction of $u_5$ and $u_1$, respectively.

We supplement the analytical calculations with numerical solutions using arc-length control.
To test the stability of each configuration, we probe each equilibrium with unit
forces along each of the displacement directions $u_4, u_5$, and $u_1$, respectively. The complete
bifurcation diagrams are shown in Fig.~\ref{fig:fccOrthorhombic} and the solution path can be
interpreted as follows. Starting from the stress free configuration ($\Theta = 0$), the temperature
of the lattice is reduced and a first instability occurs at $\Theta_1^s = -0.00188$, making the $u_4
= 0$ branch unstable thus causing the lattice to snap either in positive or negative $Y$ direction as
shown in Fig.~\ref{fig:fccOrthoa}. In real applications, small defects would cause a
unit cell to snap into one of the stable branches of the pitchfork of
Fig.~\ref{fig:fccOrthoa}. However, numerically we can continue to reduce the temperature while tracing the
unstable branch of $u_4$. In this way we can obtain the other stable branches of the pitchfork in directions
$Z$ and $X$, respectively, as show in Figs.~\ref{fig:fccOrthob} and~\ref{fig:fccOrthoc}. Finally,
Figs.~\ref{fig:fccOrthorhombic} (d)-(f) show the bifurcation diagrams in the axial directions.

\begin{figure}[ht!]  \centering
\subfigure[\label{fig:fcccubica}]{\includegraphics[width=0.32\textwidth]{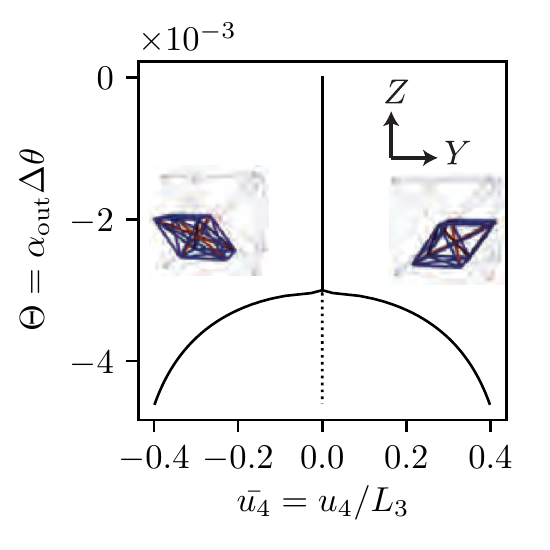} \centering}
\subfigure[\label{fig:fcccubicb}]{\includegraphics[width=0.32\textwidth]{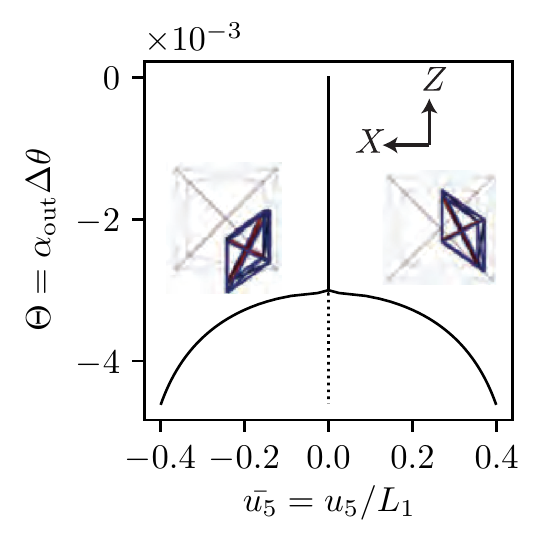} \centering}
\subfigure[\label{fig:fcccubicc}]{\includegraphics[width=0.32\textwidth]{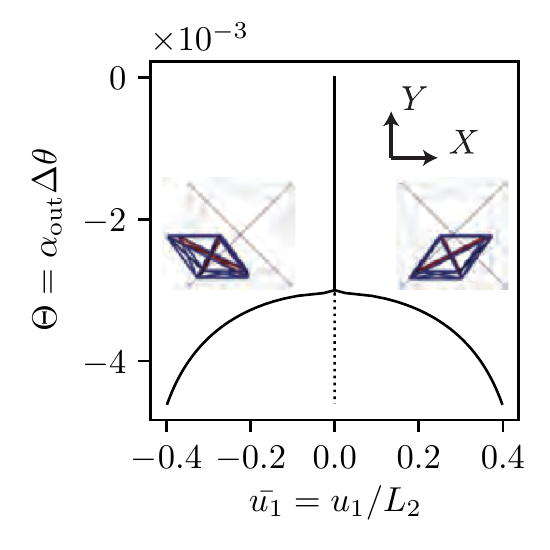} \centering}
\subfigure[\label{fig:fcccubicd}]{\includegraphics[width=0.32\textwidth]{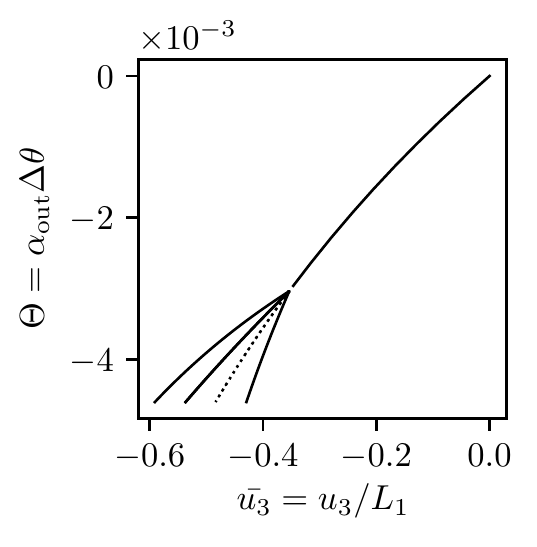} \centering}
\subfigure[\label{fig:fcccubice}]{\includegraphics[width=0.32\textwidth]{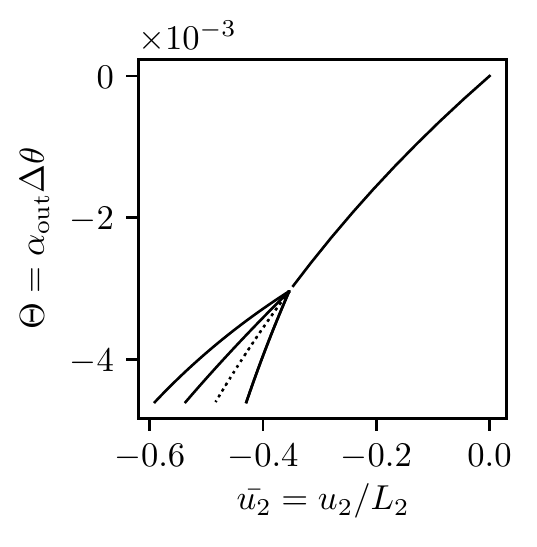} \centering}
\subfigure[\label{fig:fcccubicf}]{\includegraphics[width=0.32\textwidth]{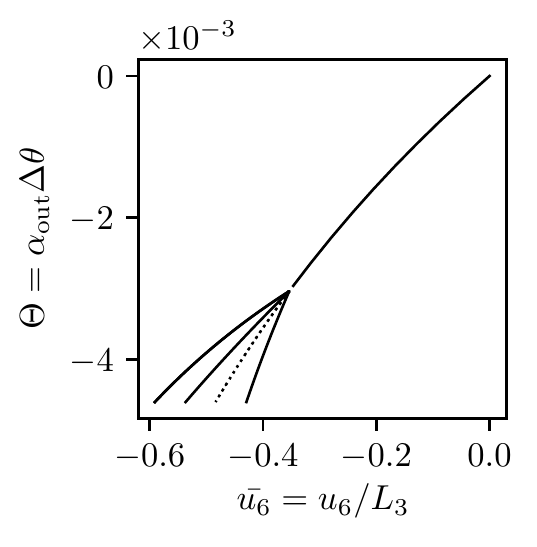} \centering}
\caption{\label{fig:fcccubic} Bifurcation diagrams for cubic fcc type lattices with
$\dfrac{L_2}{L_1} = \dfrac{L_3}{L_1} = 1$, $\dfrac{\kin}{\kout} = 1.0$,
$\dfrac{\alphain}{\alphaout} = 1000$. Due to the symmetry of the cube, the pitchfork in the three
directions $Y, Z$ and $X$ happens at the same temperature as shown in (a), (b) and (c).  (d), (e)
and (f) plot the bifurcation diagrams of the axial displacements. Insets in (a), (b) and (c) show
the bifurcated lattice in its stable branches of their respective pitchforks. Solid lines are stable
solutions while dotted lines are unstable solutions.}
\end{figure}

\subsection{Cubic fcc lattices ($L_1 = L_2 = L_3$)}
\label{subsec:cubicfcc}

We consider next the fcc-type lattice with all sides of equal length (i.~e. $L_1 = L_2 = L_3$)
stiffness ratio $\dfrac{\kin}{\kout} = 1.0$ and conductivity ratio $\dfrac{\alphain}{\alphaout} =
1000$. In this case, the lattice exhibits full symmetry. Thus as we reduce the
temperature~$\Theta$, precisely when $\Theta_s = -0.00303$, three eigenvalues of the Jacobian
$\mcl{L}(\mbs{u}, \Theta) = D_1\mbs{g}(\mbs{u}, \Theta)$ simultaneously vanish. In
\ref{subsec:Liapunov-schmidt} we described the Liapunov-Schmidt reduction when only one of the
eigenvalues becomes zero, but the procedure can be extended for problems with eigenvalues of higher
multiplicity. Here, however, we only present numerical results and summarize them in
Fig.~\ref{fig:fcccubic}. In this case, and due to the aforementioned symmetries, the pitchforks
corresponding to all singularities are identical (see Figs. \ref{fig:fcccubic}(a)-(c)). Hence, when
the singular point is crossed, all six configurations of the buckled lattices are equally favorable.

\begin{figure}[ht!]  \centering
\subfigure[\label{fig:bcccubica}]{\includegraphics[width=0.32\textwidth]{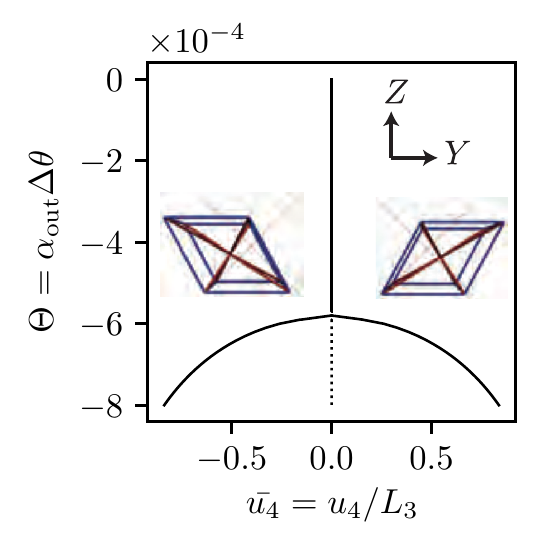} \centering}
\subfigure[\label{fig:bcccubicb}]{\includegraphics[width=0.32\textwidth]{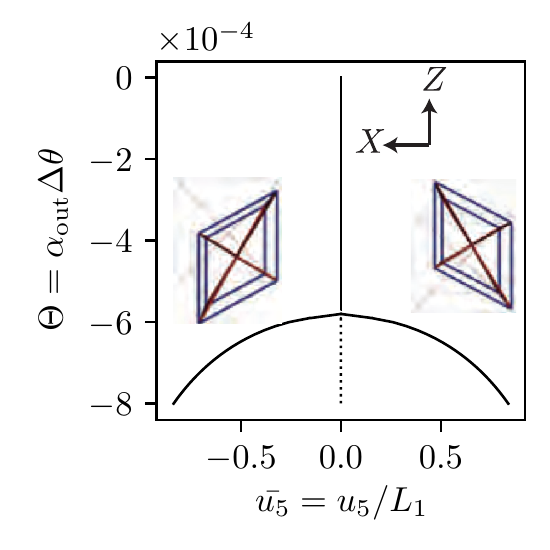} \centering}
\subfigure[\label{fig:bcccubicc}]{\includegraphics[width=0.32\textwidth]{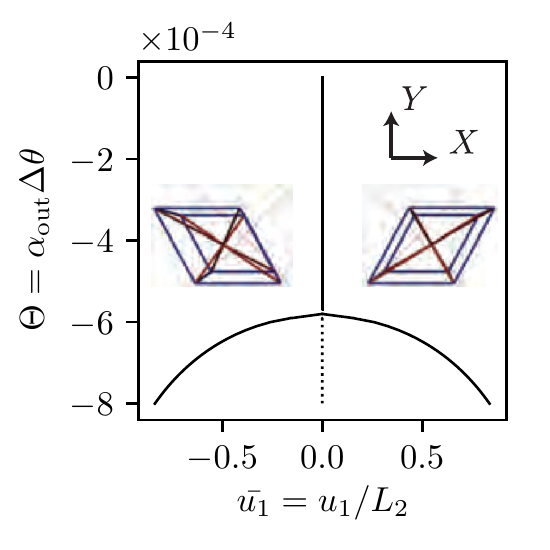} \centering}
\subfigure[\label{fig:bcccubicd}]{\includegraphics[width=0.32\textwidth]{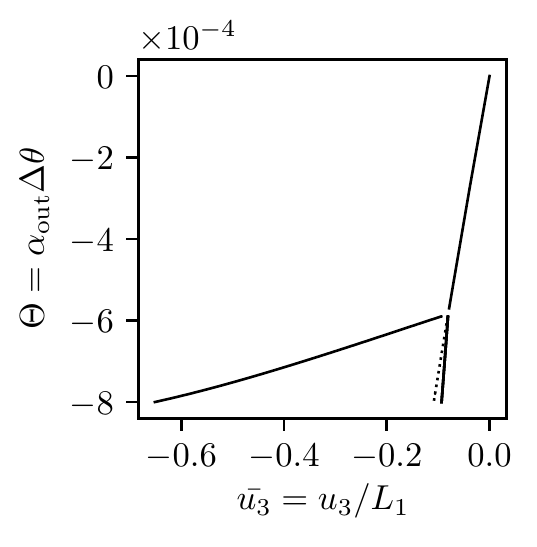} \centering}
\subfigure[\label{fig:bcccubice}]{\includegraphics[width=0.32\textwidth]{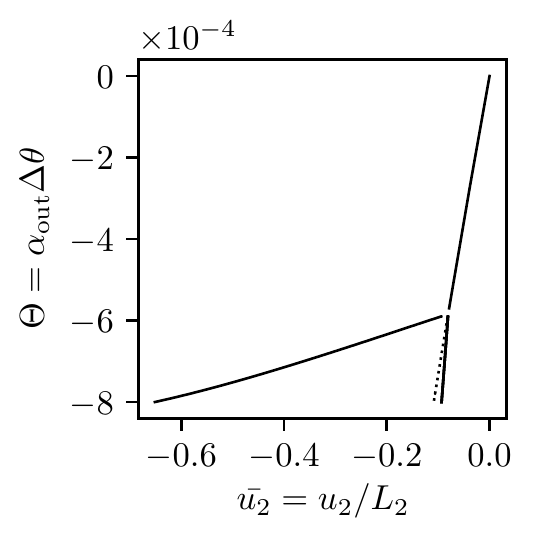} \centering}
\subfigure[\label{fig:bcccubicf}]{\includegraphics[width=0.32\textwidth]{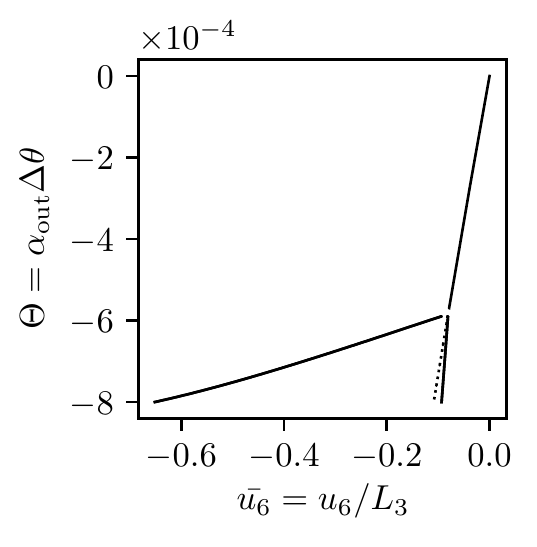} \centering}
\caption{\label{fig:bcccubic} Bifurcation diagrams for cubic bcc type lattices with
$\dfrac{L_2}{L_1} = 1$, $\dfrac{L_3}{L_1} = 1$, $\kin/\kout = 0.5$, $\alphain/\alphaout = 1000$. The
pitchforks in the three directions $Y, Z$ and $X$  all happen at the same temperature as
shown in (a), (b) and (c). (d), (e) and (f) plot the
bifurcation diagrams of the axial displacements. Insets in (a), (b) and (c) show the bifurcated
lattice in its stable branches of their respective pitchforks. Solid lines are stable solutions
while dotted lines are unstable solutions.}
\end{figure}

\subsection{Cubic bcc lattices ($L_1 = L_2 = L_3$)}

We consider bcc-type lattices that have springs only along the edges and the body diagonals
(see Fig. \ref{fig:3d-unit-cell}(b)). We first consider a symmetric geometry $L_1 = L_2 = L_3$ and
$\dfrac{\kin}{\kout} = 0.5$, $\dfrac{\alphain}{\alphaout} = 1000$. As temperature is reduced, three
of the eigenvalues of the Jacobian vanish simultaneously at $\Theta_s = -0.00058$, corresponding to
an inverted pitchfork in the three directions as shown in Figs.~\ref{fig:bcccubica}-(c).

\subsection{Orthorhombic bcc lattices  ($ L_1 \neq L_2 \neq L_3$)}
\label{subsec:orthobcc}

\begin{figure}[ht!]  \centering
\subfigure[\label{fig:bccorthoa}]{\includegraphics[width=0.32\textwidth]{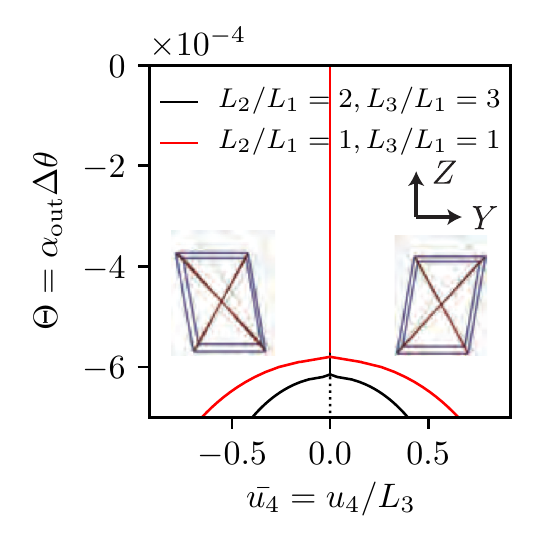} \centering}
\subfigure[\label{fig:bccorthob}]{\includegraphics[width=0.32\textwidth]{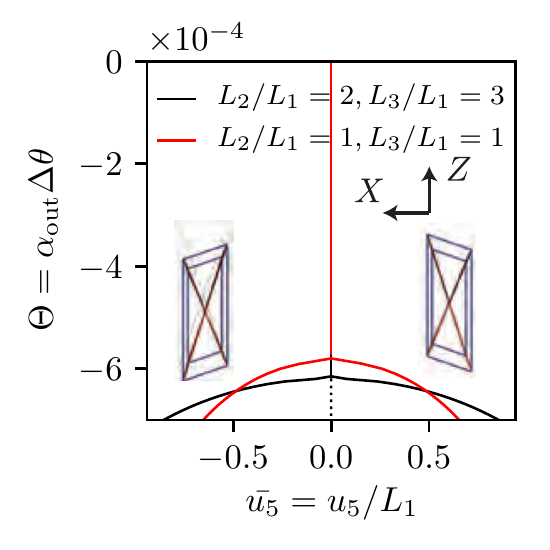} \centering}
\subfigure[\label{fig:bccorthoc}]{\includegraphics[width=0.32\textwidth]{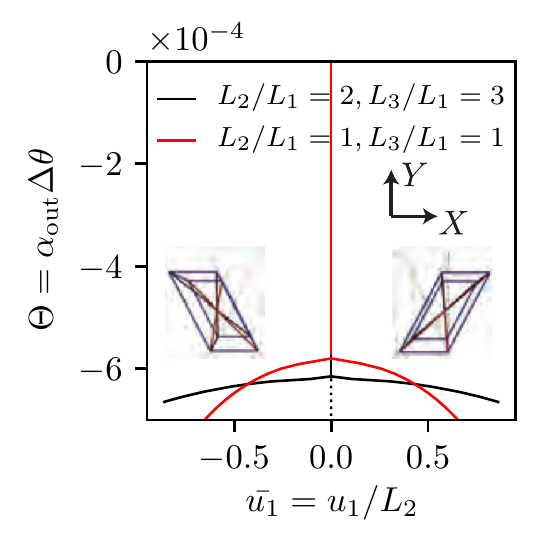} \centering}
\subfigure[\label{fig:bccorthod}]{\includegraphics[width=0.32\textwidth]{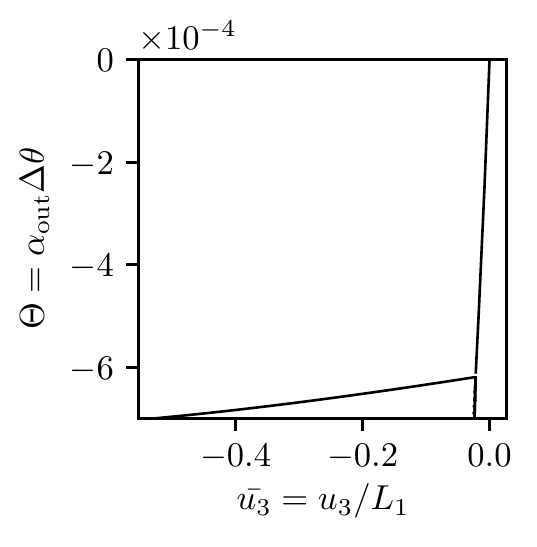} \centering}
\subfigure[\label{fig:bccorthoe}]{\includegraphics[width=0.32\textwidth]{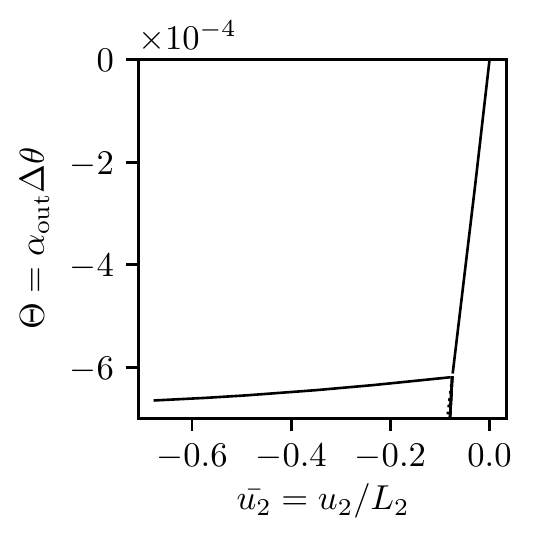} \centering}
\subfigure[\label{fig:bccorthof}]{\includegraphics[width=0.32\textwidth]{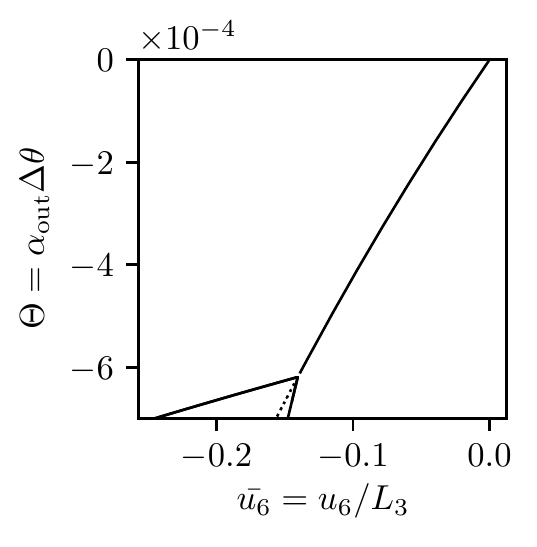} \centering}
\caption{\label{fig:bccortho} Bifurcation diagrams for orthorhombic bcc type lattices with
$\dfrac{L_2}{L_1} = 2$, $\dfrac{L_3}{L_1} = 3$, $\dfrac{\kin}{\kout} = 0.5$,
$\dfrac{\alphain}{\alphaout} = 1000$. The pitchforks in the three directions $Y, Z$ and $X$  all
take place at the same temperature despite symmetry being broken as shown in (a), (b) and
(c). However, the solutions of the pitchfork are not identical in all the three directions.  Shown
in red in (a),(b) and (c) is the pitchfork of the cubic bcc lattice ($\dfrac{L_2}{L_1} = 1$,
$\dfrac{L_3}{L_1} = 1$) with same stiffness and coefficients of thermal expansion for
comparison. (d), (e) and (f) plot the bifurcation diagrams of the axial displacements.}
\end{figure}

Finally, we analyze bcc lattices with the same material parameters as the last case but breaking the
geometric symmetry. Remarkably, even when ($L_1 \neq L_2 \neq L_3$) all three instabilities still
happen at the same temperature. In this case, the singularity point does not change, although the
solutions can considerably differ from those in the symmetric configuration far from the
bifurcation (see Figs. \ref{fig:bccorthoa}-(c)). A direct analogy can be made between this case and the two-dimensional analysis where
changing the aspect ratio of the frame did not change the bifurcation temperature (see
Section~\ref{subsec:perturbation-analysis}).

\subsection{Systematic variation of structural parameters}

\begin{figure}[ht!]  \centering
\subfigure[\label{fig:phasediagfcc}]{\includegraphics[width=0.48\textwidth]{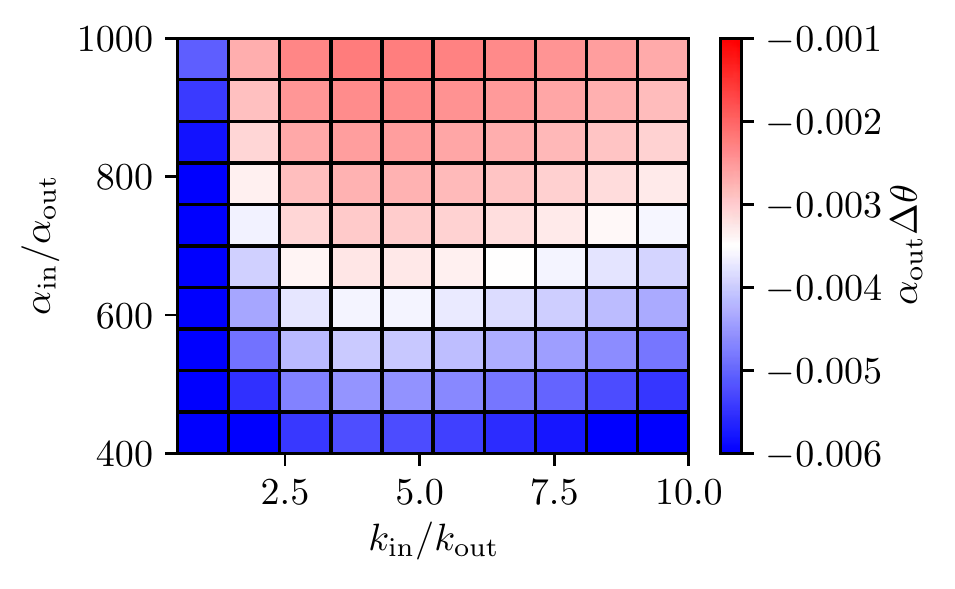} \centering}
\subfigure[\label{fig:phasediagbcc}]{\includegraphics[width=0.48\textwidth]{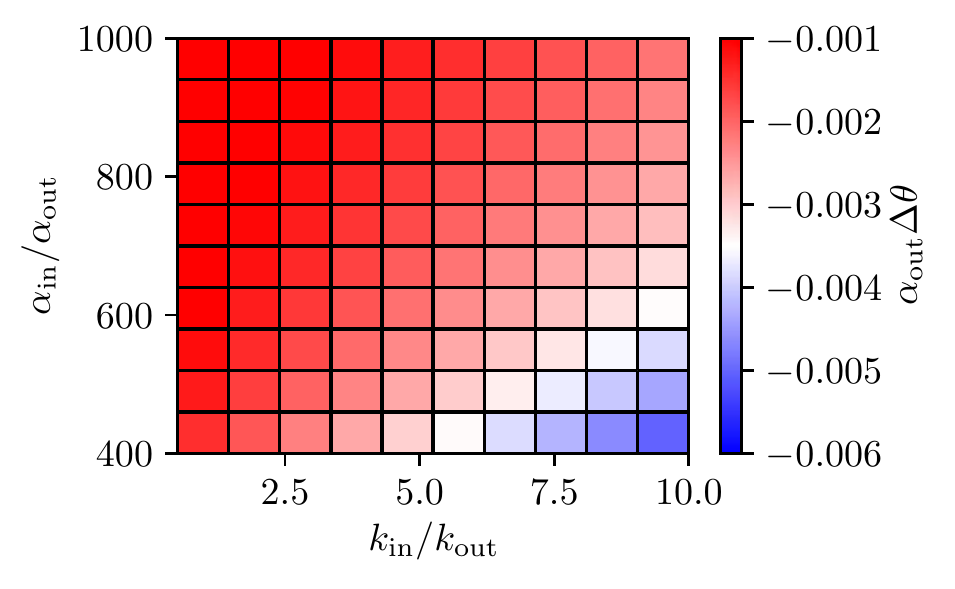} \centering}
\caption{\label{fig:phasediag-3D} Phase diagram for (a) fcc type cubic lattices with $L_1 = L_2 = L_3$
  and (b) bcc type cubic lattices with
$L_1 = L_2 = L_3$ with the colorbar representing the bifurcation temperature.}
\end{figure}

We finally study the value of the bifurcation temperature as a function of the two structural
parameters of the lattice viz. $\dfrac{\kin}{\kout}$ and $\dfrac{\alphain}{\alphaout}$.
Fig.~\ref{fig:phasediag-3D} summarizes, in the form of a phase diagram, the results obtained
for a bcc and an fcc lattice, respectively, both with $L_1 = L_2 = L_3$.

The differences between the two types of lattices are noteworthy. To justify this claim,
we note that the bottom-left corners of these two phase fields are the most interesting
regions of these plot since they refer to lattices where the differences in
stiffness and thermal expansion coefficients between the two types of springs are smallest. By
comparing Figs.~\ref{fig:phasediag-3D}(a) and (b) we conclude that bcc-type lattices
in this region buckle at higher temperature than their fcc counterparts. In practice, this
means that the temperature difference triggering the buckling of the bcc lattice is,
approximately, six times smaller than in the fcc structure.

\section{Effective (macroscopic) behavior of 3D lattices}
\label{sec:3D-macroscopic}

We have already shown in Section~\ref{sec:numerical-examples} that a lattice of $20\times20$ unit
cells had a structural response and stability behavior dictated, to a certain extent, by the
characteristics of the unit cell (see Fig. \ref{fig:complete-heat-cycle}), although boundary effects
are noticeable. Motivated again by our interest on the analysis of complex lattices, we
study next a \emph{periodic} three-dimensional lattice of $N_x \times N_y \times
N_z$ unit cells. Let~$\mathcal{B}$ be the periodic macroscopic lattice and $\partial \mathcal{B}$
denote its boundary.  Periodic boundary conditions are imposed, first,  by  partitioning
$\partial\mathcal{B}$ into two disjoint regions $\partial \mathcal{B}^+$ and $\partial \mathcal{B}^-$. Then, pairs of node positions $\mbs{X}_q^+$ and $\mbs{X}_q^-$
are identified on $\partial\mathcal{B}^+$ and $\partial \mathcal{B}^-$, respectively, such that
the periodicity constraint
\begin{equation}
    \mbs{u}_q^+ - \mbs{u}_q^- = (\mbs{F}-\mbs{I})(\mbs{X}_q^+ - \mbs{X}_q^-)
     \label{eq:pbc}
\end{equation}
is imposed \citep{Miehe2002}. In this equation $\mbs{F}$ is the macroscopic deformation
gradient. This constraint is readily implemented using Lagrange multipliers. In practice, for a
regular geometry such as a cube, a master node $M_i (i = 1,2,3)$ can be defined for each of the pair
of opposite faces and the far-field deformation gradient can be imposed on the periodic lattice
through the displacement on each of the master nodes according to
\begin{equation}
    \mbs{u}(M_i) = (\mbs{F} - \mbs{I}) \mbs{l}_i\ ,
\end{equation}
where $\mbs{l}_i = l_i\, \mbs{e_i}$ are orthogonal vectors on the axes of the cube
(see, e.g., \citep{Segurado2002}).

\begin{figure}[ht!]%
\centering%
\includegraphics[width=\textwidth]{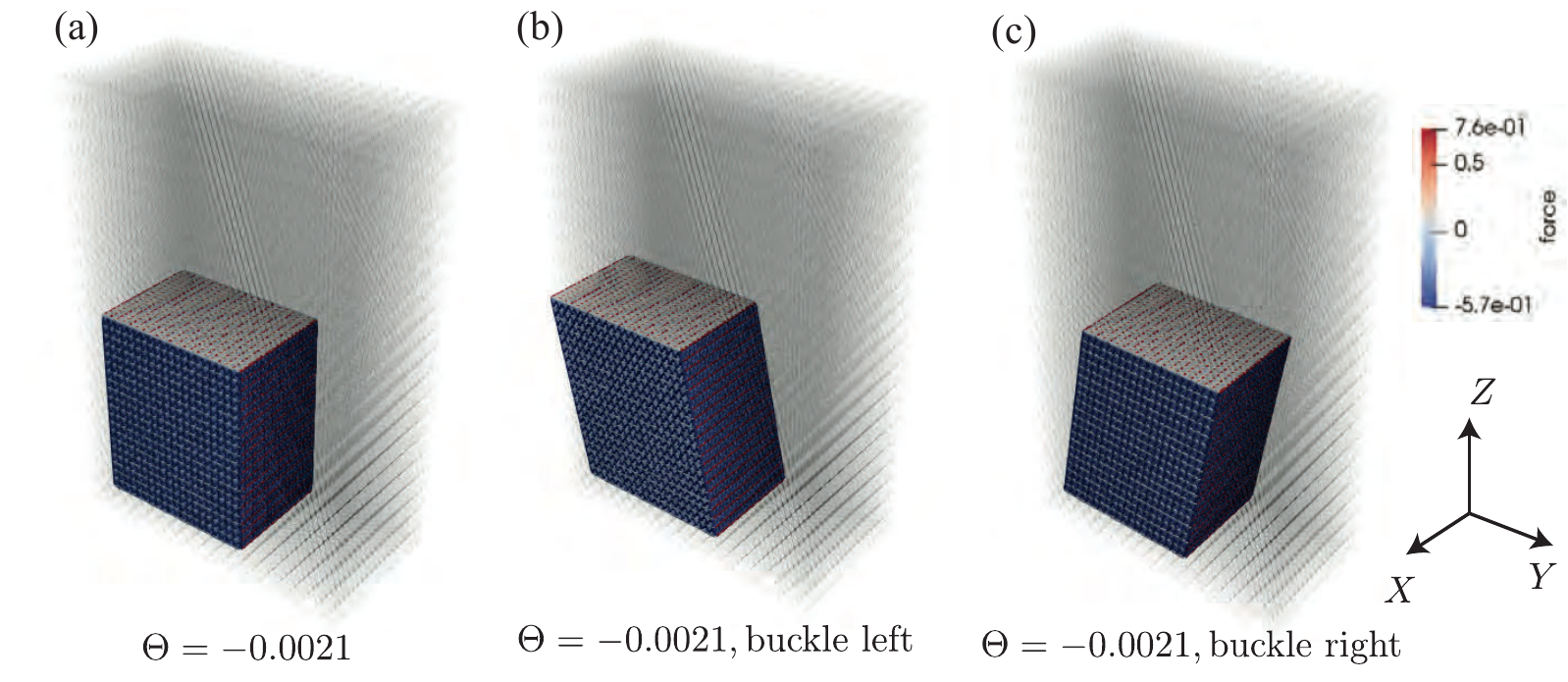}%
\caption{Buckled configurations of a $20 \times 20 \times 20$ periodic lattice comprising of
orthorhombic fcc type unit cells. (a) is the lattice configuration at $\Theta = -0.0021 $ beyond the
bifurcation point at $\Theta_1^s = -0.00188$ and is unstable. On application of a small amount of
force it buckles to either (b) or (c). The properties of the unit cell are $\dfrac{L_2}{L_1} = 2$,
$\dfrac{L_3}{L_1} = 3$, $\dfrac{\kin}{\kout} = 1$ and $\dfrac{\alphain}{\alphaout} = 1000$. }%
 \label{fig:macro-fcc}%
\end{figure}

\begin{figure}[ht!]%
\centering%
\includegraphics[width=\textwidth]{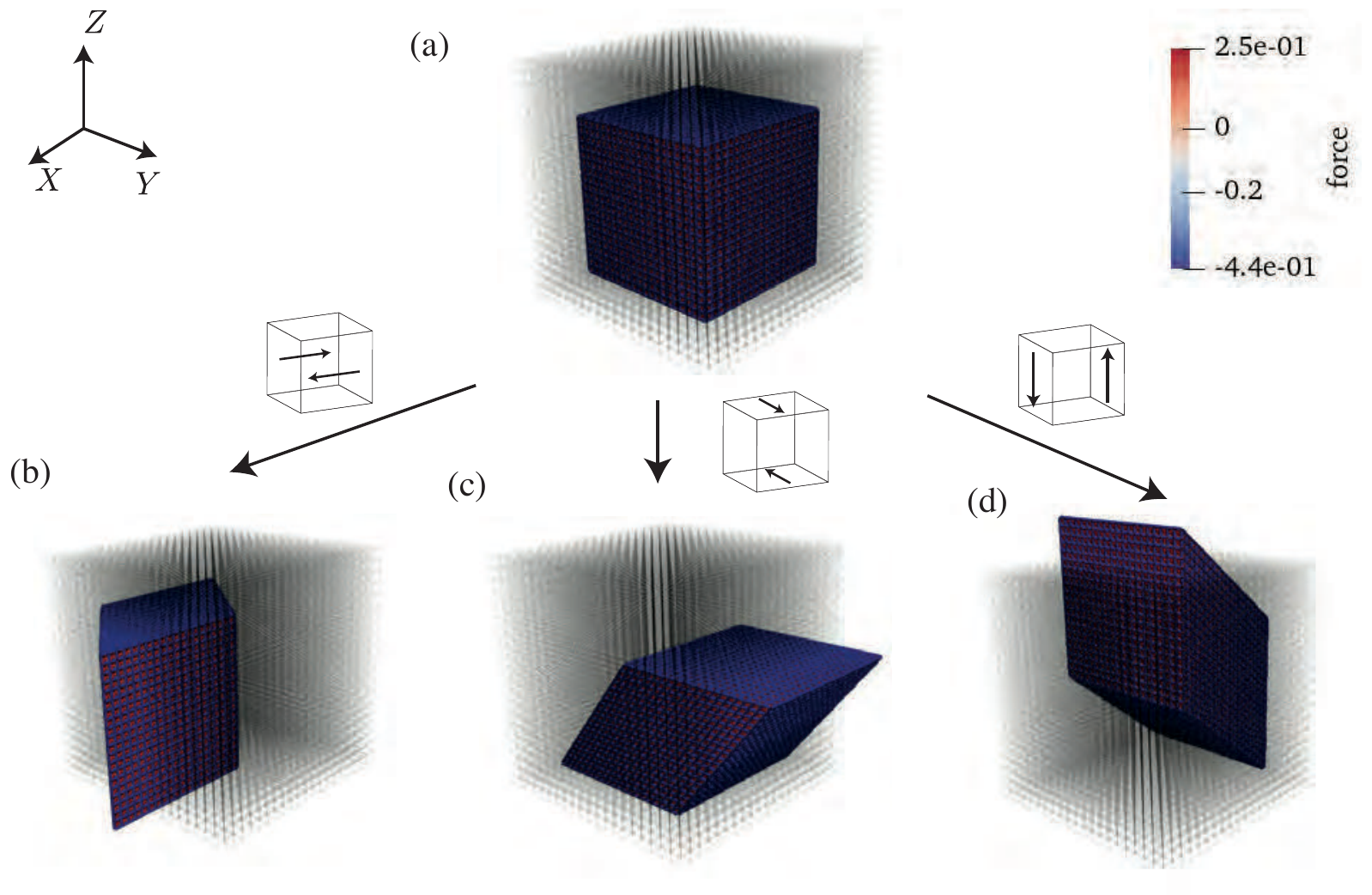}%
\caption{Buckled configurations of a $20 \times 20 \times 20$ periodic lattice comprising of cubic bcc
type unit cells. (a) is the lattice configuration at $\Theta = -0.0009 $ beyond the bifurcation point
at $\Theta^s = -0.00058$ and is unstable. As this unit cell buckles symmetrically to any of the
three directions below the bifurcation point, on a small application of force, it buckles to (b),
(c), or (d). The properties of the unit cell are $L_1 = L_2 = L_3$, $\dfrac{\kin}{\kout} = 0.5$ and
$\dfrac{\alphain}{\alphaout} = 1000$.}%
 \label{fig:macro-bcc}%
\end{figure}

\begin{figure}[ht!]%
\centering%
\includegraphics[width=\textwidth]{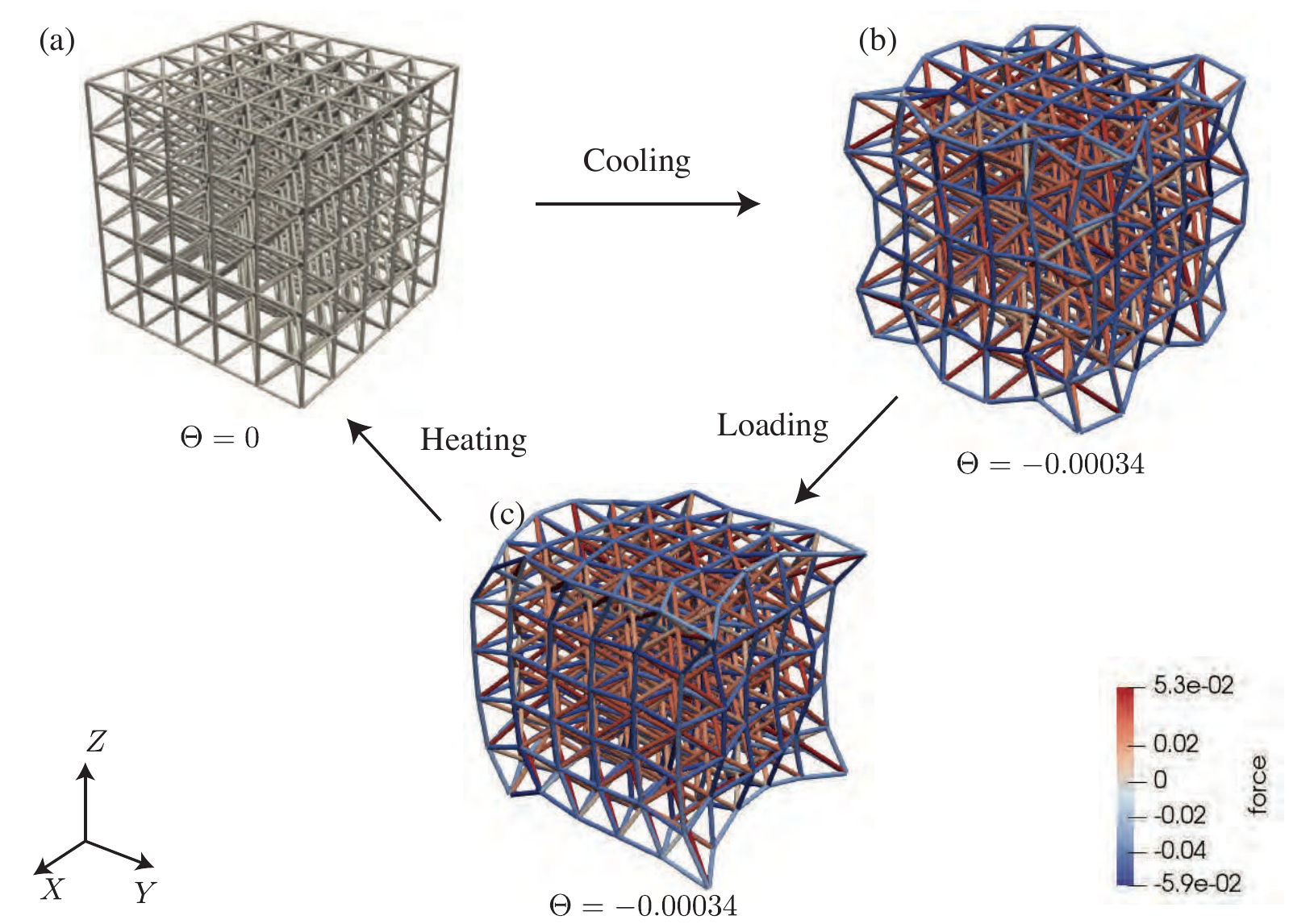}%
\caption{Complete thermomechanical cycle of a practical lattice with a Gaussian noise on the
  stiffness (with zero mean and standard deviation $0.01 \%$). The lattice consists of cubic bcc-type unit cells which can buckle symmetrically
in all three directions below the bifurcation point. (a) is the undeformed configuration at $\Theta
= 0$. (b) is the deformed configuration at $\Theta = -0.00034$, where each of the unit cells has
buckled randomly into one of the directions. (c) is the deformed configuration after a mechanical
load on (b) which on heating gives back configuration~(a).}%
 \label{fig:macro-gaussian-noise}%
\end{figure}

In Figs.~\ref{fig:macro-fcc}, \ref{fig:macro-bcc}, and \ref{fig:macro-gaussian-noise}, we
present numerical examples of large lattices built from different unit cell types and
material parameters.  We note that,
especially in those situations where structural instabilities arise,  there is a
dependency of the critical buckling parameter with the size of the
representative volume element (RVE) \citep{Herrnbock2022}. In our
numerical simulations, we drive the bifurcation parameter $\Theta$ well beyond
the critical value of the unit cell so that we ensure that buckling always takes place.

First, in Fig. \ref{fig:macro-fcc} we consider a periodic $20 \times 20 \times 20$  lattice of
orthorhombic fcc-type unit cells with $\dfrac{L_2}{L_1} = 2$, $\dfrac{L_3}{L_1} = 3$,
$\dfrac{\kin}{\kout} = 1$ and $\dfrac{\alphain}{\alphaout} = 1000$ whose bifurcation diagrams are
plotted in Fig.~\ref{fig:fccOrthorhombic}. The lattice exhibits a first bifurcation in the
$Y$-direction at $\Theta_1^s$, followed by bifurcations in $Z$ and $X$ directions, respectively, at
$\Theta_2^s$ and $\Theta_3^s$. The temperature is reduced from $\Theta = 0$ to $\Theta = -0.0021$
(far below the bifurcation point at $\Theta_1^s = -0.00188$) where the lattice is still in its
non-sheared configuration (see Fig.~\ref{fig:macro-fcc}(a)) which is unstable and on an application
of a small force buckles to one side or the other (see Figs.~\ref{fig:macro-fcc}(b), (c)). Note
that, if the temperature is further reduced for this type of unit cell, the lattice can buckle in
the other two directions.

Next, in Fig.~\ref{fig:macro-bcc} we consider a periodic $20 \times 20 \times 20$ lattice consisting of
cubic bcc-type unit cells with $L_1 = L_2 = L_3 $, $\dfrac{\kin}{\kout} = 1$ and
$\dfrac{\alphain}{\alphaout} = 1000$ whose bifurcation diagrams are plotted in
Fig~\ref{fig:bcccubic}. For this unit cell, below the critical $\Theta_s$, bifurcations in
all three directions are simultaneously possible and this behavior is shown for our macroscopic
lattice in Figs. \ref{fig:macro-bcc} (b), (c) and (d).

Finally, since defects are unavoidable in real lattices, we simulate a cubic bcc-type lattice where
a Gaussian noise is introduced in the stiffness of all the bars. Because of the presence of defects,
each of the unit cells buckles slightly below the bifurcation point and can buckle randomly in any
of the three directions. The complete thermomechanical cycle is shown in
Fig.~\ref{fig:macro-gaussian-noise}. Starting from a regular lattice depicted in
Fig.~\ref{fig:macro-gaussian-noise}(a),  after a reduction of the temperature, some cells buckle in
unpredictable directions as shown in Fig.~\ref{fig:macro-gaussian-noise}(b). On the application of a
shear load, some of the energetically-equivalent buckled configurations are favored and the
macroscopic shape of the lattice becomes stable at a tilted configuration, even when the load is
removed (see Fig.~\ref{fig:macro-gaussian-noise}(c)). Finally, when the temperature is raised to the
original value, the lattice recovers its undeformed, symmetric configuration.

This complete cycle confirms the theoretical possibility of building lattice metamaterials that can
absorb energy (in the low temperature phases) and be healed by simply heating them up. This is the
main outcome of the methodology developed in this paper.


\section{Concluding remarks}
\label{sec:discussion}
In this article, we have employed singularity theory to guide the design of lattice structures which
display bistable and thermally reversible behavior, so that they mimic at the macro-scale the
characteristic thermomechanical coupling of shape-memory alloys. In particular, we have investigated
structures that possess several energetically equivalent configurations at ``low'' temperature, all
of them accessible when external loading is applied, and a single stable phase at ``high''
temperature.

Singularity theory has enabled to characterize the stability of these structures at all temperatures
and served to identify the main physical material-related parameters that affect the response of the
lattice. A key outcome of this work is that the analytical predictions obtained from singularity
theory for unit cells have been compared with numerical simulations of  large scale two- and
three-dimensional lattices, showing that the bifurcation behaviour of the unit cell determines the
macroscopic response of the structures.  To the authors' knowledge, this is the first article that
provides a complete structural analysis based on singularity theory of the stability of thermally
reversible metamaterials inspired by shape-memory alloys.

One additional outcome of singularity theory is the identification of \emph{all} the lattices
perturbations that can significantly modify their bifurcation diagrams, and hence their
stability. Since no real lattice is perfect due to, e.g., manufacturing defects and material
uncertainties, the precise characterisation of critical imperfections should be important for
guiding inspection campaigns and design constraints. While such practical aspects are not covered in
this work, the methodology presented can be useful for these related problems.

The investigation presented in this paper has, as long-term goal, the design of energy absorbing
structures that can be healed with a heat treatment after, for example, an impact. A final comment
must be added regarding the possibility of manufacturing real lattices with the proposed
features. For the specific lattice geometry investigated in this article, its physical realization
is limited by the materials involved in the two types of springs (internal and outer frame
springs). As shown in Section~\ref{sec:2D-unit-cell}, the thermal expansion coefficients of these
two elements must differ by a factor of $600$ so as to provide the lattice with the required
bistability and thermal reversibility features in the two-dimensional case. For three-dimensional
geometries, however, the ratio is only around $400$. These ratios severely restrict the material
pairs that could be employed to actually manufacture such lattices. The main result of the article
is hence, not the specific lattice design, but the methodology presented, which remains useful to
study other lattice configurations in the search of shape-memory metamaterials.


\appendix

\section{Singularity theory approach to bifurcation problems}
\label{sec:singularity-theory}

Singularity theory mainly deals with the study of the solution set in equations of the form
\begin{equation}
  g(x,\lambda) = 0
  \label{Eq:normForm}
\end{equation}
where $g:\mathbb{R}\times\mathbb{R}\to\mathbb{R}$, $x$ is the \emph{state variable} and $\lambda$ is the \emph{bifurcation parameter}. Bifurcation problems deal specifically with equations where the number solutions~$x$ that solve Eq.~\eqref{Eq:normForm} changes with~$\lambda$. The graph of pairs $(x,\lambda)$ that satisfy Eq.~\eqref{Eq:normForm} is called the \emph{bifurcation diagram} of the system.

In the simplest problem, and in what follows, the function $g(x,\lambda)$ is infinitely differentiable. The stability and bifurcation analysis starts by identifying \emph{singular points} of the solution
\begin{equation}
  g(x_0,\lambda_0) = 0 \quad \mathrm{and} \quad g_{,x} (x_0,\lambda_0) = 0\ ,
  \label{eq-singular}
\end{equation}
where $g_{,\alpha}$ refers, here and below, to the partial derivative $\partial_\alpha g$. If $g_{,x}(x_0,\lambda_0) \neq 0$, sufficiently close to $(x_0,\lambda_0)$, the state variable $x$ of the solution set can be expressed as a function of $\lambda$ and hence there cannot be a change in the number of solutions near $\lambda=\lambda_0$. In order to have a bifurcation at $(x_0,\lambda_0)$ it is thus necessary, although not sufficient, that this point be singular.

\begin{figure}[t]
\centering
\includegraphics[width=0.5\textwidth]{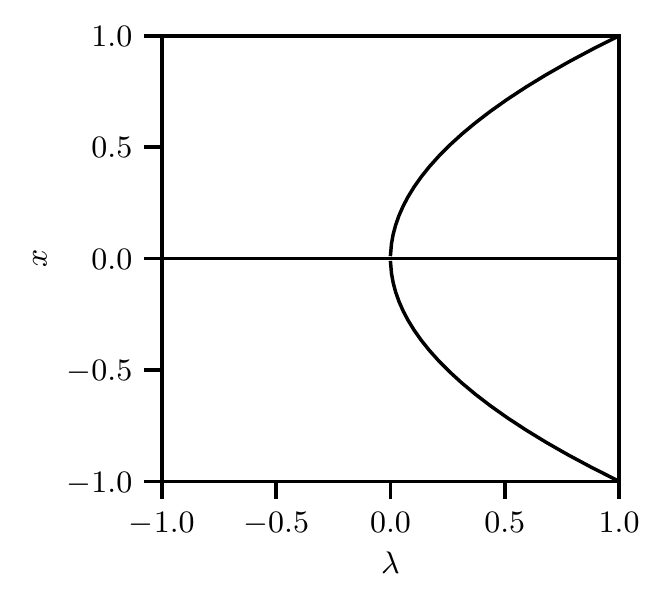}
\caption{Bifurcation diagram for  $g(x,\lambda) = x^3 - \lambda x = 0$. }
\label{fig:fig_2}
\end{figure}

Many types of bifurcations can occur, in general, but the only relevant type for our analysis is the
\emph{pitchfork bifurcation} (see Figs. \ref{fig:2D_diagrams}-\ref{fig:fig_7}). Its defining feature
is that the number of solutions changes from one to three when the bifurcation parameter crosses a
critical value. The so-called \emph{normal form} of the pitchfork bifurcation is
\begin{equation}
  g(x,\lambda) = x^3 - \lambda x\ .
  \label{eq-pitchfork}
\end{equation}
The number of solutions that satisfy $g(x,\lambda)=0$ changes from one to three as $\lambda$ varies from negative to positive values (see Fig.~\ref{fig:fig_2} for an illustration of the corresponding bifurcation diagram).

The first central issue that singularity theory addresses is the identification of the necessary conditions under which a certain bifurcation problem is qualitatively equivalent to that of a simpler norm form. This is referred to as the \emph{recognition problem}.

\subsection{The recognition problem}
\label{subsec:recogprob}

The recognition problem refers to the identification of the nature of a bifurcation using only
information from the derivatives of~$g$ at the point of singularity. If Eq.~\eqref{Eq:normForm} has
a singularity at $(x_0,\lambda_0)$ and the conditions
\begin{equation}
g = g_{,x} = g_{,xx} = g_{,\lambda} = 0\;, \qquad g_{,xxx}\, g_{,\lambda x} < 0
\label{Eq:pitchfork-conditions}
\end{equation}
are satisfied at $(x_0,\lambda_0)$, then $n(\lambda)$, the number of solutions of $g(x,\lambda)=0$ expressed as a function of the bifurcation parameter, changes from one to three when $\lambda$ changes from small negative values to small positive values, and the problem exhibits a pitchfork bifurcation. Similarly, if the last condition is replaced by $g_{,xxx}g_{,\lambda x} > 0$, then $n(\lambda)$ changes from three to one and it will correspond to an inverted pitchfork.

Moreover, the theory tells that if conditions in Eq.~\eqref{Eq:pitchfork-conditions} are satisfied, then $g(x,\lambda)$ is \emph{equivalent} to the normal form of the pitchfork bifurcation~\eqref{eq-pitchfork} in the sense that there exists a nonlinear transformation of coordinates $(X(x,\lambda),\Lambda(\lambda))$ such that

\begin{enumerate}
\item It is a local diffeomorphism of $\mathbb{R}^2$ of the form $(x,\lambda) \rightarrow (X(x,\lambda),\Lambda(\lambda))$ mapping the origin to~$(x_0,\lambda_0)$.
\item A nonzero function $ S(x,\lambda)$ exists such that
 \begin{equation}
   S(x,\lambda) g(X(x,\lambda),\Lambda(\lambda)) = x^3 - \lambda x\ .
   \label{Eq:equivalence}
 \end{equation}
\end{enumerate}

Since the factor $S(x,\lambda)$ is nonzero, the solutions of $g(x,\lambda) = 0$ differ from those of $x^3 - \lambda x = 0$ only by the diffeomorphism $(X,\Lambda)$.

%

\subsection{Influence of parameters: universal unfoldings}
\label{subsec:unfoldings}

To define the influence of perturbations on the solution set of a given function~$g$, let $\alpha =
\{\alpha_1, \alpha_2, ..., \alpha_k \}$ be a set of $k$ auxiliary parameters. Then, let us construct
a certain family of functions $G(x,\lambda,\alpha)$ such that
\begin{equation}
  G(x,\lambda, 0) = g(x,\lambda)\ .
  \label{eq-perturbation}
\end{equation}
The function $G(x,\lambda,\alpha)$ is called a \emph{perturbation} of $g$. For a given bifurcation problem, singularity theory identifies the minimum set of auxiliary parameters that contains all possible perturbations of $g$, called the \emph{universal unfolding} of~$g$. For the pitchfork bifurcation, two auxiliary parameters are required to construct its universal unfolding
\begin{equation}
  G(x,\lambda,\alpha_1,\alpha_2) = x^3 - \lambda\, x + \alpha_1 + \alpha_2\, x^2\ .
  \label{Eq:universal-unfolding}
\end{equation}

\begin{figure}[t]
\centering
\includegraphics[width=0.7\textwidth]{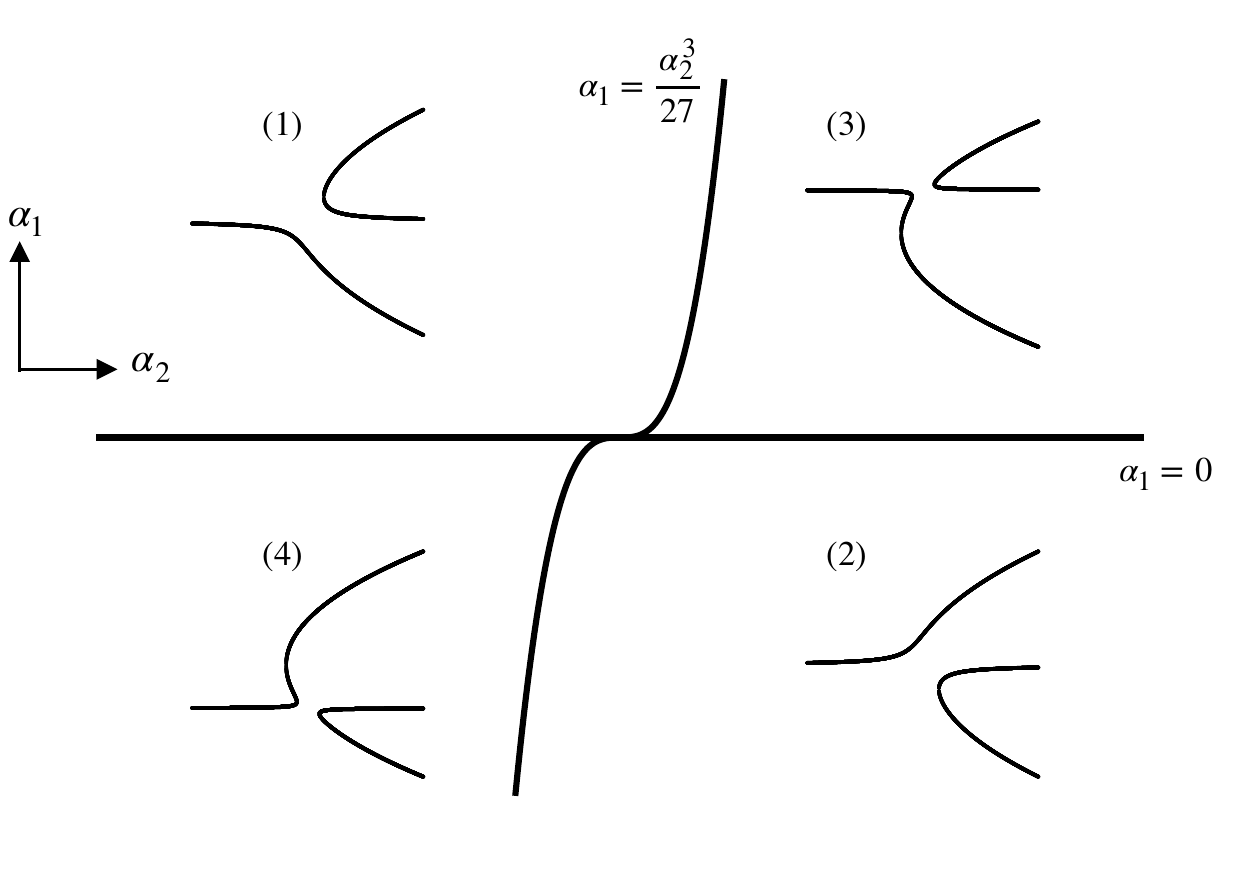}
\caption{Universal unfolding of a pitchfork bifurcation: Regions (1) and (2) are usually
perturbations that occur only by a single parameter, while the kinked bifurcation diagrams in (3) and
(4) are obtained only under specific imperfection values.}
\label{fig:fig_3}
\end{figure}

Thus, the universal unfolding~\eqref{Eq:universal-unfolding} contains all possible perturbations
of~\eqref{eq-pitchfork}. Moreover, singularity theory calculates how the bifurcation diagram depends
on the auxiliary parameters $\alpha_1,\alpha_2$ (see Fig.~\ref{fig:fig_3} for the pitchfork
bifurcation). The theory predicts four regions in the $(\alpha_1,\alpha_2)$ plane that correspond to
qualitatively different perturbations of the pitchfork. The central result is that these are the
only possible perturbations for the pitchfork.

In real applications, imperfections may occur in infinitely many ways, but all imperfections can be ``lumped together'' into two auxiliary parameters that cannot generate bifurcation diagrams that are qualitatively different from the four types mentioned above. The main result that allows this identification is as follows. Let $G(x,\lambda,\alpha_1, \alpha_2)$ be any two-parameter unfolding of $g(x,\lambda)$. Then, $G$ is a \emph{universal unfolding} of~$g$ if and only if
\begin{equation}
  \rm{det}
  \begin{pmatrix}
    0 & 0  & g_{,x\lambda} & g_{,xxx}\\
    0 & g_{,\lambda x}  & g_{,\lambda \lambda} & g_{,\lambda x x} \\
    G_{,\alpha_1} & G_{,\alpha_1 x}  & G_{,\alpha_1 \lambda} & G_{,\alpha_1 x x} \\
    G_{,\alpha_2} & G_{,\alpha_2 x}  & G_{,\alpha_2 \lambda} & G_{,\alpha_2 x x} \\
  \end{pmatrix} \neq 0
  \label{Eq:RecogProblem-unfolding}
\end{equation}
at $ x = \lambda = \alpha_1 = \alpha_2 = 0$. This is called the \emph{recognition problem for the universal unfolding}.


\subsection{Liapunov-Schmidt reduction}
\label{subsec:Liapunov-schmidt}

We have presented heretofore singularity theory for equations where the state variable $x$ is a
scalar. In real applications, however, $x$ often belongs to a space of multiple or even infinite
dimensions. The \emph{Liapunov-Schimdt reduction} can reduce a finite or infinite dimensional system
of equations to one of a single dimension to which singularity theory can then be applied. Let us
consider now an equilibrium equation of the form
\begin{equation}
   \mbs{g}(\mbs{x},\mbs{\alpha}) = \mbs{0}\ ,
   \label{Eq:equlibrium-equation-ndim}
\end{equation}
where $\mbs{g}: \mathbb{R}^n \times \mathbb{R}^{k+1} \to \mathbb{R}^n$, with $n>1$ is a smooth mapping,
$\mbs{x} = (x_1,x_2,\cdot \cdot \cdot,x_n)$  are the unknowns and $\mbs{\alpha} = (\alpha_0,
\alpha_1,\cdot \cdot \cdot,\alpha_k)$ are auxiliary parameters where $\alpha_0$ can be considered to
be the bifurcation parameter. Without loss of generality, we assume that $\mbs{g}(\mbs{0},\mbs{0})
= \mbs{0}$ and define
\begin{equation}
  \mcl{L} = D_{1}\mbs{g}(\mbs{0},\mbs{0})
  = \left.\dfrac{\partial g_i}{\partial x_j}\right|_{(\mbs{0},\mbs{0})}
\label{eq-jacobian}
\end{equation}
to be the Jacobian matrix. If the Jacobian has full rank, i.e., $\mathrm{rank}(\mcl{L}) = n$, then
by the implicit function theorem $\mbs{x}$ can be uniquely solved as a function of $\mbs{\alpha}$.

Here, we are interested in problems where $\mcl{L}$ is singular and we focus next on the minimally
degenerate case, where the Jacobian has a simple zero eigenvalue and thus $\mathrm{rank} (\mcl{L}) = n-1$.
The reduction process begins by choosing vector spaces $\mcl{M}$ and $\mcl{N}$
that complement the kernel and the range of $\mcl{L}$, respectively, i.e., $\mathbb{R}^n = \mathrm{ker}~\mcl{L}\oplus \mcl{M}  = \mcl{N} \oplus \mathrm{range} ~\mcl{L}$. Since $\mathrm{rank}(\mcl{L}) = n-1$, the dimension of $\mathrm{ker}~\mcl{L}$ is one and
$\mathrm{dim}~\mathrm{range} ~\mcl{L} = n-1 $  which gives $ \mathrm{dim}~\mathcal{M} = n-1$ and
$\mathrm{dim}~\mathcal{N} = 1$. Next, let us define a projection $E:
\mathbb{R}^n\to\mathrm{range}~\mcl{L}$.  The system of equations
Eq.~\eqref{Eq:equlibrium-equation-ndim} can then be split into an equivalent pair of equations,
namely,
\begin{subequations}
  \begin{align}
      E\mbs{g}(\mbs{x},\mbs{\alpha}) &= \mbs{0}\ , \label{Eq:eqsplita}\\
      (I-E)\mbs{g}(\mbs{x},\mbs{\alpha}) &= \mbs{0}\ , \label{Eq:eqsplitb}
   \end{align}
\end{subequations}
where $I$ is the identity matrix. Expressing any $\mbs{x} \in \mathbb{R}^n$ as $\mbs{x} = \mbs{v} + \mbs{w}$,  with  $\mbs{v} \in\mathrm{ker}~\mcl{L}$ and $\mbs{w}\in\mathcal{M}$, we rewrite Eq.~\eqref{Eq:eqsplita} as $ \mathcal{F}(\mbs{v},\mbs{w},\mbs{\alpha}) = \mbs{0}$ with $\mathcal{F}(\mbs{v},\mbs{w},\mbs{\alpha}) := E\mbs{g}(\mbs{v}+\mbs{w},\mbs{\alpha})$, where $\mathcal{F} : \mathrm{ker}~\mcl{L}\times \mcl{M}\times \mathbb{R}^{k+1} \to \mathrm{range}~\mcl{L}$. Taking the derivative of the function $\mathcal{F}$ with respect to the variable $\mbs{w}$, we have that $D_{2}\mathcal{F} = E D_{1} \mbs{g} = E \mcl{L} =: \mcl{\bar{L}}$ and $\mcl{\bar{L}} : \mathcal{M} \times \mathbb{R}^{k+1} \to \mathrm{range}~\mcl{L}$ is invertible. Hence, $\mbs{w}$ can be written uniquely in terms of $\mbs{v} $ and $\mbs{\alpha}$, thus we write $\mbs{w} = \mbs{W}(\mbs{v},\mbs{\alpha})$.  The map $\mbs{W} : \mathrm{ker}~\mcl{L}\times\mathbb{R}^{k+1} \to \mcl{M}$ satisfies
\begin{equation}
  E \mbs{g}(\mbs{v} + \mbs{W}(\mbs{v},\mbs{\alpha}),\mbs{\alpha}) = 0 \hspace{12.5pt}
  \mathrm{and}\hspace{12.5pt}\mbs{W}(\mbs{0},\mbs{0}) = \mbs{0}\ .
   \label{Eq:definition-of-W}
\end{equation}
Replacing $\mbs{W}(\mbs{v},\mbs{\alpha})$ in Eq.~\eqref{Eq:eqsplitb}, we obtain the reduced
map $\phi: \mathrm{ker}~\mcl{L} \times \mathbb{R}^{k+1} \to \mcl{N}$ defined as
\begin{equation}
   \phi(\mbs{v},\mbs{\alpha}) = (I-E)\mbs{g}(\mbs{v}+\mbs{W}(\mbs{v},\mbs{\alpha}),\mbs{\alpha})\ .
   \label{Eq:reducedEq}
\end{equation}

The solutions to $\phi(\mbs{v},\mbs{\alpha}) = 0$ are in one-to-one correspondence with the
solutions of Eq.~\eqref{Eq:equlibrium-equation-ndim}, and thus it is the reduced equation we are
looking for. By introducing coordinates for the subspaces $\mathrm{ker}~\mcl{L}$,
$\mathrm{range}~\mcl{L} $, $\mcl{M}$,  $\mcl{N}$, Eq.~\eqref{Eq:reducedEq} can be further simplified
to depend only on real numbers. To see this, let $\mbs{v_0}$ and $\mbs{v_0}^*$ be arbitrary nonzero
vectors in $\mathrm{ker}~\mcl{L}$ and $(\mathrm{range}~\mcl{L})^{\perp}$, respectively. Any vector
$\mbs{v} \in \rm{ker}~\mcl{L}$ can be uniquely written as $\mbs{v} = x \mbs{v_0}$, with
$x\in\mathbb{R}$. We define $g:\mathbb{R}\times\mathbb{R}^{k+1}\to\mathbb{R}$ by
\begin{equation}
   g(x,\mbs{\alpha}) = \langle \mbs{v_0}^*,\phi(x\mbs{v_0},\mbs{\alpha}) \rangle\ ,
   \label{Eq:reducedg}
\end{equation}
where $\langle\cdot,\cdot\rangle$ refers to the standard Euclidean inner product.
The roots of Eq.~\eqref{Eq:reducedg} are the same as those of Eq.~\eqref{Eq:reducedEq} and  hence
also to the ones of the original Eq.~\eqref{Eq:equlibrium-equation-ndim}. Substituting
Eq.~\eqref{Eq:reducedEq} in  Eq.~\eqref{Eq:reducedg}, the operator $ (I-E) $ drops out leading to
the final reduced equation:
\begin{equation}
  g(x,\mbs{\alpha}) = \langle \mbs{v_0}^*,\mbs{g}(x\mbs{v_0} +
\mbs{W}(x\mbs{v_0},\mbs{\alpha}),\mbs{\alpha}) \rangle\ .
   \label{Eq:reduced-final}
 \end{equation}

\subsection{Recognition conditions for the reduced equation}

To investigate the bifurcation behavior of the reduced equation, the partial derivatives of $g(x,\mbs{\alpha})$ need to systematically obtained using the chain rule. For that, we introduce an invariant notation for the higher order derivatives of a function of several variables. If $(\mbs{v_1},\cdot \cdot \cdot, \mbs{v_n}) \in \mathbb{R}^n$, we define
\begin{equation}
   (D^k_{1}\mbs{g})_{\mbs{x},\mbs{\alpha}} (\mbs{v_1}, \cdot \cdot \cdot , \mbs{v_k}) = \pd{}{t_1} \cdot \cdot \cdot \pd{}{t_k} \mbs{g} \left.\left( \mbs{x} + \sum_{i=1}^{k} t_i\mbs{v_i}, \mbs{\alpha}\right)\right|_{t_1 = \cdot \cdot \cdot = t_k = 0}\ .
\end{equation}

Using this definition, the formulae for the first few derivatives of $g$ evaluated at $(\mbs{0}, \mbs{0})$ are
\begin{subequations}
   \begin{align}
      g_{,x} &= 0\ , \label{Eq:LSeqa}  \\
      g_{,xx} &= \langle \mbs{v_0^*} ~, ~D^2_{1}\mbs{g} (\mbs{v_0},\mbs{v_0}) \rangle\  \label{Eq:LSeqb} \\
     g_{,xxx} &= \langle \mbs{v_0^*}~, ~ D^3_{1}\mbs{g} (\mbs{v_0},\mbs{v_0}, \mbs{v_0}) + 3D^2_{1}\mbs{g}(\mbs{v_0},\mbs{W}_{,xx})\rangle
                \ ,\label{Eq:LSeqc}\\
      g_{,\alpha_l} &= \langle \mbs{v_0^*}~, ~ \mbs{g}_{,\alpha_l}\rangle\ , \label{Eq:LSeqd}\\
      g_{,\alpha_l x}&= \langle \mbs{v_0^*}~, ~ D_{1}\mbs{g}_{,\alpha_l} \cdot \mbs{v_0} +  D^2_{1}\mbs{g}(\mbs{v_0}, \mbs{W}_{,\alpha_l})\rangle\ , \label{Eq:LSeqe}\\
      g_{,\alpha_m \alpha_l}&=\langle \mbs{v_0^*}~, ~ \mbs{g}_{,\alpha_m \alpha_l} + D_{1}\mbs{g}_{,\alpha_m} \mbs{W}_{,\alpha_l} + D_{1}\mbs{g}_{,\alpha_l} \mbs{W}_{,\alpha_m} + D^2_{1}\mbs{g}(\mbs{W}_{,\alpha_m},\mbs{W}_{,\alpha_l})\rangle\ , \label{Eq:LSeqf}\\
      \begin{split} \label{Eq:LSeqg}
          g_{,\alpha_l x x } &= \langle \mbs{v_0^*}~, ~ D^2_{1}\mbs{g}_{,\alpha_l} (\mbs{v_0}, \mbs{v_0}) + D^3_{1}\mbs{g}(\mbs{v_0},\mbs{v_0},\mbs{W}_{,\alpha_l}) + 2 D^2_{1} \mbs{g} (\mbs{v_0}, \mbs{W}_{,\alpha_l x}) \\
          & \hspace{30pt} D_{1}\mbs{g}_{,\alpha_l} (\mbs{W}_{,xx}) + D^2_{1}\mbs{g}( \mbs{W}_{,\alpha_l}, \mbs{W}_{,xx})\rangle\ .
      \end{split}
   \end{align}
\end{subequations}

These derivatives depend themselves on the derivatives of $\mbs{W}(x\mbs{v_0},\mbs{\alpha})$ that can be obtained by differentiating Eq.~\eqref{Eq:definition-of-W} implicitly with respect to $\mbs{x} $ or $\mbs{\alpha}$.

\begin{acknowledgements}
  \myack
\end{acknowledgements}

\bibliographystyle{\mybibstyle}
\bibliography{driver}
\end{document}